\documentclass{IEEEtran} 
\pdfoutput=1
\usepackage{header}

\title{Uncertainty in Multi-Commodity Routing Networks: When does it help?
}

\author{Shreyas Sekar, Liyuan Zheng, Lillian J. Ratliff, and Baosen Zhang
\thanks{All of the authors are associated with the Department of Electrical Engineering, University of Washington, Seattle, WA 98195-2500, USA.
        (email: {\tt\small \{sekarshr, liyuanz8, ratliffl, zhangbao\}@uw.edu}).}
        \thanks{This work was supported by NSF awards CNS-1634136 and CNS-1646912.}
}

\begin{document}

\maketitle
\thispagestyle{empty}
\pagestyle{empty}

\begin{abstract}
We study the equilibrium behavior in a multi-commodity selfish routing game with
many types of uncertain users where
each user over- or
under-estimates their congestion costs by a multiplicative factor. Surprisingly,
we find that uncertainties in different directions have qualitatively distinct
impacts on  equilibria. Namely, contrary to the usual notion that uncertainty
increases inefficiencies, network congestion actually decreases when users over-estimate their
costs. On the other hand, under-estimation
of costs leads to increased congestion.
We apply these results to urban transportation networks, where drivers have different estimates about the cost of congestion.
In
light of the dynamic pricing policies aimed at tackling congestion, our results indicate that users' perception of these prices can significantly impact the policy's efficacy, and ``caution in the face of uncertainty'' leads to favorable network conditions.
\end{abstract}

\begin{IEEEkeywords}
	Network Routing, Uncertainty, Nash Equilibrium, Transportation
\end{IEEEkeywords}

\section{Introduction}
\label{sec:intro}
\IEEEPARstart{M}{ulti-commodity} routing networks that allocate resources to self-interested users lie at the heart of many systems such as communication,
transportation, and power networks (see, e.g.,~\cite{ParekhEtAl2014} for an
overview). In all of these systems,  users \emph{inherently face uncertainty and
are heterogeneous}. These users rarely have perfect information about the state
of the system, and each have their own idiosyncratic objectives and tradeoffs
between time, money, and risk~\cite{stidham04,stupiaSBV15,azouziA03}. Naturally, these
private beliefs influence each user's decision and, in turn, the total welfare of the overall system.
In this paper, we provide an understanding of the effects of certain classes of
uncertainties and user heterogeneity with respect to such uncertainties
on  network performance. In particular, we establish conditions on when they are helpful and harmful to the overall social welfare of the system.

A motivating example of a routing network that we use throughout this paper is the \emph{urban transportation network}.
Commuters in road networks simultaneously
trade-off between diverse objectives such as total travel time, road taxes,
parking costs, waiting delays, walking distance and environmental impact. At the
same time, these users tend to possess varying levels of information and
heterogeneous attitudes, and there is
evidence~\cite{devetag2008playing,gao2011cognitive} to suggest that the routes
adopted depend not on the true costs but on how they are perceived by the users.
For instance, users prefer more consistent routes over those with high variance~\cite{chen2003travel}, seek to minimize travel time over parking costs~\cite{pierce2013getting}, and react adversely to per-mile road taxes~\cite{schaller2010new}.

Furthermore, the technological and economic incentives employed by planners interact with user beliefs in a complex manner~\cite{brownM17}.
For example, to limit the economic losses arising from urban congestion, cities
across the world have introduced a number of solutions including road
taxes, time-of-day-pricing,  road-side message signs and route recommendations~\cite{economist,schrank2015,mortazavi2009travel,kricheneRAB14}. However, the dynamic nature of these
incentives (e.g., frequent price updates) and the limited availability of
information dispersal mechanisms may add to users' uncertainties and asymmetries
in beliefs.





The effect of uncertainties on network equilibria has been examined in recent
work~\cite{brownM17,thaiLB16,wuLA17,tavafoghi2017informational} where  
each user perceives the network condition to be different than the true conditions. 
The current results have mostly focused on simple network topologies (e.g.,
parallel links) or networks where a fixed percentage of the
population is endowed with a specific level of uncertainty. Given the complexity of most practical networks, it is natural to ask how
uncertainty~(i.e.~user beliefs) affects equilibria in scenarios with many types
of users, whose perceptions vary according to the user type. Specifically, in
this work we answer the following two questions: (i) \emph{how do
    equilibria depend on the type and level of uncertainty in
networks with multiple types of users}, and (ii) \emph{when does uncertainty improve or degrade equilibrium quality?}

To address these questions, we turn to a \emph{multi-commodity selfish routing} framework commonly employed by many disciplines
(see, e.g.,
\cite{Korilis97,acemoglu2016informational,farokhiKBJ13,qiu2003selfish}).
In our model,  each user seeks to route some flow from a source to a destination across a network and faces congestion costs on each link. These congestion costs are perceived differently by each user in the network, representing the uncertainties in their beliefs.

It is well-known that even in the  presence of perfect information~(every user
knows the exact true cost), strategic behavior by the users can result in
considerably worse congestion at equilibrium when compared to a centrally
optimized
routing solution~\cite{roughgardenT02}.
Against this backdrop, we
analyze what happens when users have imperfect knowledge of the congestion
costs. A surprising outcome arises: {in the presence of uncertainty, if users
    select routes based on perceived costs that over-estimate the true cost,
    \emph{the equilibrium quality is better compared
    to perfect information case}. Conversely, if the users are not cautious and under-estimate the costs, the equilibrium quality becomes worse.

	\subsection{Contributions}
 We introduce the notion of \emph{type-dependent uncertainty} in multi-commodity
 routing networks, where the uncertainty of users belonging to type $\theta$ is
 captured by a single parameter $r_\type > 0$. Specifically, for each user of
 type $\theta$, if their true cost on edge $e$ is given by $C_e(x) = a_e x +
 b_e$, where $x$ is the total population of users on this edge, then their \emph{perceived cost} is $r_\theta a_e x +  b_e$.



	For the majority of this work, we will focus on \emph{cautious behavior}, where users over-estimate
 the costs  ($r_\theta \geq 1$), for all types $\theta$. Some of our results will also hold for the case where users under-estimate
	the costs ($r_{\theta} \leq 1$), for all types $\theta$. The central message
    of this paper is that when users exhibit ``caution in
    the face of uncertainty'', the social cost at the equilibrium is smaller
    compared to the case where users have perfect information (i.e.~know the
    true congestion costs).
 
    The following results are \emph{independent of network topology}:
    \begin{enumerate}
		\item[(a)] The social cost---i.e.~$\C(\vec{x}) =\sum_{e\in \E} x_eC_e(x_e)$
where $x_e$ is the total population mass, summed over all user types, flowing on edge
$e$---of the equilibrium\footnote{For the multi-commodity routing game studied in this work, the equilibrium is essentially unique as all equilibria induce the same flow on the edges; see Proposition~\ref{prop:potential}}
            solution where all users have
		the same level of uncertainty ($r_\type = r$ for all $\type$) is
		\emph{always smaller than or equal to} the cost of the equilibrium
		solution without uncertainty when $r \in [1,2]$ and vice-versa
        when $r < 1$.
		\item[(b)] The worst-case ratio of the social cost of the equilibrium to
		that of the socially optimal solution (i.e.~the price of
		anarchy~\cite{roughgardenT02}) is $4/(4\rmax\gamma - \rmax^2)$,
		where $\rmax = \max_\type r_\type$ and $\gamma \leq 1$ is the ratio of the minimum to the maximum uncertainty over user types.
		\end{enumerate}

Constraining network topology, we show the following:
		\begin{enumerate}
			\item[(a)] The social cost of the equilibrium where a fraction of
                the users exhibit an uncertainty of $r \in [1,2]$ and the
                rest have no uncertainty \emph{is always smaller} than or equal
                to the social cost of an identical system without uncertainty,
                as long as the network has the serially linearly independent topology (a subclass of series-parallel networks~\cite{milchtaich06}).

				\item[(b)] In systems having users with and without uncertainty,
		the routing choices adopted by the uncertain users always results in an
		improvement in the costs experienced by users without uncertainty, as long as the graph has a series-parallel topology~\cite{milchtaich06}.

	\end{enumerate}
	At a high level, the main contribution of this work can be viewed as a
    nuanced characterization of the instances (i.e.~network topologies, levels
    of uncertainty) for which uncertainty is helpful in reducing the
    equilibrium congestion levels. We show that these characterizations
    are tight by means of illustrative examples where uncertainty leads to an increase in the social cost when our characterization conditions are violated. Finally, many of our results also extend gracefully to instances having polynomial edge cost functions, which are of the form $C(x) = ax^d + b$ (see Section~\ref{sec:polynomial}).

	To validate the theoretical results, we present a number of simulation results.
	We focus specifically on the application of parking in urban transportation
	networks and consider realistic urban network topologies with two types of
	users: through traffic and parking users.
    Given a parking population with uncertainties, we show that \emph{cautious behavior improves equilibrium quality while lack of caution degrades it} even when uncertainty is asymmetric across user types and when the same user faces  different levels of uncertainty on different parts of network. 	
    \subsection{Comparison with Other Models of Uncertainty}
	Our work is closely related to the extensive body of work on risk-averse selfish
	routing~\cite{fotakisKL15,nikolovaS15,Altman02} and pricing tolls in congestion
	networks~\cite{stidham04,brownM17,jelinekKS14}. The former line of research
	focuses on the well known \emph{mean-standard deviation} model where each self-interested
	user selects a path that minimizes a linear combination of their expected travel
	time and standard deviation. While such an objective is desirable from a central
	planner's perspective, experimental studies suggest that individuals tend to
	employ simpler heuristics when faced with uncertainty~\cite{tversky1975judgment}.
	Motivated by this, we adopt a multiplicative model of uncertainty similar to~\cite{mavronicolasMMT07,meirP15,Qi17}.

The literature on
	computing tolls for heterogeneous users is driven by the need to
	\emph{implement the optimum routing} by adjusting the toll amount, which is often
	interpreted as the \emph{time--money} tradeoff, on each edge. It is possible to draw parallels
	between additive uncertainty models
	(such as the mean--standard deviation model) and tolling; specifically, tolling
	can be viewed as adjusting for users' uncertainty.
	While tolls can be (within reason) arbitrarily decided, the system planner has little
    influence over the level of uncertainty among the users. Bearing this in
    mind, we strive for a more subtle understanding of how equilibrium congestion depends on the level of uncertainty.
Moreover, while many of the results in this paper focus on the effect of
over-estimating costs, different than the existing literature, we also study
the effect of cost under-estimation.


	Finally, our work is thematically similar to the recent paper on the \emph{informational Braess paradox}~\cite{acemoglu2016informational}
    whose model can be viewed as an extreme case of our model where the
    uncertainty parameter $r_{\type}$ is so high on some edges ($r_{\type}
    \rightarrow \infty$) that users always avoid such edges. On the other hand,
    our model is more continuous as user attitudes are parameterized by a finite
    value of $r_{\type}$, which allows for a more realistic depiction of the
    trade-offs faced by users who must balance travel time, congestion, and
    uncertainty. Although both the characterizations make use of the
    serially linearly independent topology,
    it is worth pointing out that our results do not follow from~\cite{acemoglu2016informational} and require different techniques that capture a more continuous trade-off between uncertainty and social cost.
	\subsection{Organization}
The rest of the paper is structured as follows. In Section~\ref{sec:model}, we formally
introduce our model followed by our main results in Sections~\ref{sec:theory} and~\ref{sec:heterogeneity}. Section~\ref{sec:simulations} presents our simulation results on urban transportation networks with parking and routing users who face different levels of uncertainty. Finally, we conclude with discussion in Section~\ref{sec:conclusions}.

\section{Model and Preliminaries}
\label{sec:model}
We consider a non-atomic, multi-commodity selfish routing game with multiple types of users. Specifically,
we consider a network represented as $G = (V,\E)$ where $V$ is the set of
nodes and $\E$ is the set of edges.  For each edge $e \in \E$, we define a
linear cost function
\begin{equation}
    C_e(x_e) = a_e x_e + b_e,
    \label{eq:ce}
\end{equation}
where $x_e \geq 0$ is the total
population (or flow) of users on that edge and $a_e, b_e \geq 0$. One can
interpret $C_e(\cdot)$ as the true cost or expected congestion felt by the users
on this edge. However, due to uncertainty, users may perceive the cost on each
edge $e \in \E$ to be different from its true cost.

To capture that users may have different perceived uncertainties, we introduce
the notion of \emph{type}. Specifically, we consider a finite set of user types
$\mathcal{T}$, where each type $\theta \in \T$ is uniquely defined by the
following tuple $(s_\theta, t_\theta, \typesize, r_\type)$. The parameter
$\typesize > 0$ denotes the total population of users belonging to type $\theta$
such that each of these infinitesimal users seeks to route \emph{some flow}
from its source node $s_\theta \in V$ to the destination node $t_\theta \in V$.
Moreover,  the parameter $r_{\theta} > 0$ captures the beliefs or uncertainties
associated with users of type $\theta$ and affects the edge cost in the
following way: users
 of type $\theta$ perceive the cost of edge $e\in \E$ to be
\begin{equation}
\label{eqn_uncertaintyperception}
\hat{C}^{\theta}_e(x_e) = r_{\theta} a_e x_e + b_e.\end{equation}

For illustration, consider an urban transportation network. Then $b_e$ may represent the
constant travel time on a link (in the absence of other vehicles) and $a_ex_e$,
the congestion-dependent component of the travel time. A multiplicative
uncertainty of $r _{\type} > 1$ indicates that users of type $\type$ adversely
view costs arising due to congestion (e.g., waiting in traffic) when compared to other costs. Alternatively, $a_ex_e$ could also represent a congestion-dependent toll or tax that is commonly levied in transportation infrastructure such as highways or parking, and the parameter $r_{\type}$ captures the time-money trade-off~\cite{fleischerJM04}.

A path $p\in
\mc{P}_\theta$ is a sequence of edges connecting $s_\type$ to $t_\type$.
Define $\mc{P}_\type$ to be the set of all $s_\type$--$t_\type$ paths in
$G$.
Let $x^\type_p\in \mb{R}$ be the
total flow routed by users of type $\type$ on path $p \in \mc{P}_{\type}$.
We use the notation $\vec{x}=(x^\type_p)_{\type\in \mc{T}, p\in \mc{P}_\type}\in
\mb{R}^{|\mc{T}|\cdot|\mc{P}_{\type}|}$ for a network flow and $\vec{x}^\theta=(x_p^\theta)_{p\in
    \mc{P}_\theta}$ to denote the network flow of type $\theta\in \mc{T}$. Then, for each type $\theta\in \mc{T}$, define the set of feasible flows to be
\begin{equation}
    \mc{X}_\theta=\{\vec{x}^\theta|\  \textstyle \sum_{p\in \mc{P}_\theta}x_p^\theta=\typesize, \ x_p^\theta\geq 0,
    \ \forall p\in \mc{P}_\theta\}.
    \label{eq:typefeasflows}
\end{equation}
%
The
action space of users of type $\theta$ is $\mc{X}_\theta$---that is, users of
type $\theta$ choose a feasible flow $\vec{x}^\theta\in \mc{X}_\theta$.
Further, define the joint action space $\mc{X}=(\mc{X}_\theta)_{\theta\in \mc{T}}$---i.e.~the space of feasible
flows for all user types.

Path flows induce edge flows. Let $x_e^\type\in \mb{R}$ be the flow on edge $e$ due to users of type $\theta$. The
edge and path flow for users of type $\theta$ are related by
\begin{equation*}
\textstyle x^\type_e = \sum_{p \in \mathcal{P }_\type, e \in p}x^\type_p.
\end{equation*}
Define the total flow on edge $e$ to be
\begin{equation*}
  \textstyle  x_e =\sum_{\type \in \mathcal{T}}x^{\type}_e.
\end{equation*}
Then, using this notation, we write the path cost in terms of edge flow. For any
path $p$,
\begin{equation}
\textstyle    C_p(\vec{x}) = \sum_{e \in p} C_e (x_e)= \sum_{e \in p}(a_e x_e + b_e).
    \label{eq:pathtoedge}
\end{equation}
Similarly, the perceived path costs are given by
\begin{equation}
\textstyle    \hat{C}_p^\theta(\vec{x})=\sum_{e\in p} \hat{C}_e^\theta(x_e)=\sum_{e\in p}
  r_\theta  a_e x_e+b_e.
    \label{eq:perceivedcostpath}
\end{equation}


Define the game instance
\begin{equation*}
    \mc{G}=\{(V,\E),\mc{T},\mc{X}, (\mc{P}_{\theta})_{\theta\in \mc{T}}, (s_\type,t_\type, \typesize, r_{\theta})_{\type\in
    \mc{T}}, (C_e)_{e\in \E}\}.
\end{equation*}
The game instance $\mc{G}$ captures all of the relevant information about the
multi-commodity routing game including the notion of type-based uncertainty we
are interested in studying.

\subsection{Nash Equilibrium Concept}

We assume that the users in the system are self-interested and route their flow
with the goal of minimizing their individual cost. Therefore, the solution
concept of interest in such a setting is that of a Nash equilibrium, where each
user type routes their flow on minimum cost paths with respect to their perceived
cost functions and the actions of the other users.

%

\begin{defn}[Nash Equilibrium]
    Given a game instance $\mc{G}$, a feasible flow $\vec{x}\in\mc{X}$ is said to be a \emph{Nash
    equilibrium} if for every $\type\in \mc{T}$, for all $p\in\mc{P}_{\type}$
    with positive flow, $x_p^\theta>0$,
\begin{equation}
\label{eq:eqn_equilibrium_inequality}
\hat{C}^{\type}_p(\vec{x}) \leq \hat{C}^{\type}_{p'}(\vec{x}),\ \ \forall \ p'\in \mc{P}_{\type}
\end{equation}
\end{defn}

For the rest of this work, we will assume that all the flows considered are feasible.

\begin{remark}[User Beliefs] For the sake of completeness and to understand how
    an equilibrium is reached, we briefly comment on each user's beliefs about
    the uncertainties of the other users.
  We assume that type-based uncertainty
    is not only known within the type but across all types---that is, a user of
    type $\theta$ knows the uncertainty levels of the users of all other types.
    While a user knowing the uncertainty level within its
    own type may not be unreasonable, full knowledge of the uncertainties of the
    other types is a strong assumption.

    This being said, for the types of games
    we consider---games admitting a potential function, which we formally
    define in Proposition~\ref{prop:potential}---a number of
    \emph{myopic learning rules}\footnote{By \emph{myopic learning rules}, we mean
    rules for iterated play that require each player to have minimal-to-no knowledge of other
players' cost functions and/or strategies.} converge to Nash equilibria (see,
e.g.,~\cite{monderer1996potential} and references therein).
We are currently investigating layered belief structures coupled with a Bayesian Nash
equilibrium concept. Some recent work has investigated such structures in a
routing game context~\cite{liu:2016aa}; however, it is well known that equilibria in these types of games are computationally
difficult to resolve.
\end{remark}

\subsection{Social Cost and Price of Anarchy}

 One of the central goals in this work is to compare the quality of the
 equilibrium solution in the presence of uncertainty to the socially optimal
 flow as, e.g., computed
by a centralized planner with the goal of minimizing the aggregate cost in the system. Specifically, the social cost of a flow $\vec{x}$ is given by
\begin{equation}
  \textstyle  \text{C}(\vec{x}) = \sum_{e \in
    \E}C_e(x_e)x_e.
    \label{eq:socialcost}
\end{equation}
Note that the social cost is only measured with respect to the true congestion costs and thus does not reflect users' beliefs or uncertainties.

To capture inefficiencies, we leverage the well-studied
notion of the \emph{price of anarchy} which is the ratio of the social cost of the \emph{worst-case Nash
equilibrium} to that of the socially optimal solution.  Formally, given an
instance $\mc{G}$ of a multi-commodity routing game belonging to some class
$\mathcal{C}$ of instances (a class refers to a set of instances that share some property) suppose that $\vec{x}^\ast$ is the flow that minimizes
the social cost $\text{C}(\vec{x})$ and that $\vec{\tilde{x}}$ is the
Nash equilibrium for the given instance, then the price of anarchy is defined as
follows.
\begin{defn}[Price of Anarchy]
    Given a class of instances $\mathcal{C}$,  the \emph{price of anarchy} for this class is
    \begin{equation}
   \textstyle \max_{\mc{G} \in \mathcal{C}}  \C(\vec{\tilde{x}})/\C(\vec{x}^\ast).
        \label{eq:poa}
    \end{equation}
\end{defn}
Since, we study a cost-minimizing game, the price of anarchy is always greater than or equal to one.

%

\section{Main Results}
\label{sec:theory}
%
\label{sec:potential}

To support these main theoretical results, we first show that our game is a
\emph{potential game}.
Routing games that fall into the general class of potential games have a number
of nice properties in terms of existence, uniqueness, and computability~\cite{monderer1996potential}.
General multi-commodity, selfish routing games with heterogeneous users,
however, do not belong to the class of potential games unless certain
assumptions on the edge cost structure are met~\cite{farokhiKBJ13}.

In our case, since we have linear latencies for each type and the uncertainty
parameter only appears on the  $a_ex_e$ term for each edge, the following proposition states that  the game instances of the form we consider admit a potential function and hence,
there always exists a Nash equilibrium~\cite{farokhiKBJ13}.
\begin{proposition}
	\label{prop:potential}
A feasible flow $\vec{x}$ is a Nash equilibrium for a given instance $\mc{G}$ of a multi-commodity routing game with uncertainty vector $(r_{\type})_{\type \in \mc{T}}$ if and only if it minimizes the following potential function:
\begin{equation}
\textstyle\Phi_r(\vec{x}) = \sum_{e \in \E}  \left(\frac{1}{2}a_ex^2_e + b_e
\sum_{\type \in \mc{T}} \frac{ 1}{r_{\type}} x^\type_e\right) \label{eq:potentialfunction}
\end{equation}
Moreover, for any two minimizers $\vec{x}, \vec{x'}$, $C_e(x_e) = C_e(x'_e)$ for every edge $e \in \E$. \end{proposition}

Note that although users perceive the multiplicative uncertainty $r_{\type}$ 	on the $a_e$ term (see Equation~\eqref{eqn_uncertaintyperception}), the parameter appears in the denominator of the $b_e$ term in the potential function above. Conceptually, these have a similar effect: dividing Equation~\eqref{eq:eqn_equilibrium_inequality} by $r_{\type}$ on both sides, one can obtain equivalent equilibrium conditions where the $r_{\type}$ term is present in the denominator of the constant $b_e$. 

\begin{IEEEproof}
    By definition, a feasible flow $\vec{x}\in\mc{X}$ is a Nash equilibrium if the following condition is satisfied for all $\theta \in \mc{T}$ and for all $p,p' \in \mc{P}_{\type}$ with $x^{\type}_p > 0$:
\[\textstyle\sum_{e \in p}(r_{\type} a_e x_e + b_e)  \leq \sum_{e \in
p'}(r_{\type} a_e x_e + b_e).\]
Since $r_{\type} > 0$, this is equivalent to $\sum_{e \in p}( a_e x_e +
\frac{b_e}{r_{\type}})  \leq \sum_{e \in p'}( a_e x_e + \frac{b_e}{r_{\type}})$.
The remainder of proof trivially follows from standard arguments pertaining to the
minimizer of a convex function. See~\cite{monderer1996potential} for more detail. 
\end{IEEEproof}
The second part of the proposition indicates that the equilibria are essentially unique as the cost on every edge is the same across solutions.

\subsection{Effect of Uncertainty on Equilibrium Quality}
Our first main result
identifies a special case of the general multi-commodity game for which
uncertainty helps improve equilibrium quality---i.e.~decreases the social
cost---whenever users over-estimate their costs by a small factor
and vice-versa when they under-estimate costs. 
To show this result, we need the following technical lemma.
\begin{lemma}
	\label{lem_technical_socialcost}
Given an instance $\mc{G}$ of a multi-commodity selfish routing game with Nash
equilibrium $\tilde{\vec{x}}=(\tilde{x}_e)_{e\in \E}$, we have that for any feasible flow
$\vec{x}$,
\begin{equation}
\textstyle    \text{C}(\tilde{\vec{x}}) -  \text{C}(\vec{x}) \leq -\sum_{\type \in \T} (\frac{2}{r_\type} - 1)\sum_{e \in \E}b_e \Delta x^{\type}_e,
\end{equation}
where $\Delta x^\type_e = \tilde{x}^{\type}_e - x^\type_e$.

\end{lemma}
The proof of the above lemma is provided in
Appendix~\ref{app:lem:technical_socialcost}.

Given an instance $\mc{G}$ of the multi-commodity routing game, we define
$\mc{G}^1$ to be the corresponding game instance with no uncertainty---that is,
$\mc{G}^1$ has the same graph, cost functions, and user types as $\mc{G}$, yet $r_\type = 1$  for all $\type \in \T$.

\begin{theorem}
	\label{thm_instancewise}
	Consider any given instance $\mc{G}$ of the multi-commodity routing game with
    Nash equilibrium $\tilde{\vec{x}}$ and the corresponding game instance
    $\mc{G}^1$, having no uncertainty, with Nash equilibrium $\vec{x}^1$. Suppose $r_\type=r$ for all $\type \in \T$.
    Then, the following hold:
	\begin{enumerate}
\item $\text{C}(\tilde{\vec{x}}) \leq \text{C}(\vec{x}^1)$ if $1 \leq r \leq 2$.
\item $\text{C}(\tilde{\vec{x}}) \geq \text{C}(\vec{x}^1)$ if $0 \leq r \leq 1$.
	\end{enumerate}
%
\end{theorem}

\emph{Remark}: What happens when the users are highly cautious, i.e., $r_{\type} > 2$ for all
$\type$? Due to the presence of a few negative examples where the social cost
increases in the presence of uncertainty, we cannot conclusively state that
uncertainty helps or hurts for all instances. However, these negative instances
appear to be isolated---both our price of anarchy result (Theorem~\ref{thm:poa})
and our experiments (Section~\ref{sec:simulations}}) validate our claim that
caution in the face of uncertainty helps the users by lowering equilibrium
social costs even when $r_{\type} > 2$---i.e., uncertainty is favorable when the users are very cautious. It is however, interesting to note that although under-estimation always leads to a worse equilibrium, over-estimation may lead to better or worse equilibria. 
\begin{IEEEproof}[Proof of Theorem~\ref{thm_instancewise}]
	Let  $\Phi_r(\vec{x})$ denote the potential function for the instance
    $\mc{G}$ and $\Phi_1(\vec{x})$ denote the potential function for
    $\mc{G}^1$ where $\Phi_1$ is given in \eqref{eq:potentialfunction} with
    $r = 1$. By definition of the potential function, we
    know that $\Phi_r(\tilde{\vec{x}}) - \Phi_r(\vec{x}^1) \leq 0$ and
    $\Phi_1(\vec{x}^1) - \Phi_1(\tilde{\vec{x}}) \leq 0$. Expanding these two inequalities, we get that
	\begin{align*}
	\textstyle	\sum_{e \in \E}\left(\frac{a_e\tilde{x}^2_e }{2}+ b_e \sum_{\type \in \mc{T}} \frac{ \tilde{x}^\type_e}{r} - \frac{a_e(x^1_e)^2}{2} - b_e \sum_{\type \in \mc{T}} \frac{x^{1,\type}_e}{r}\right) \leq 0,
        	\end{align*}
and
\begin{align*}\textstyle
	\sum_{e \in \E}\left(\frac{a_e(x^1_e)^2}{2} + b_e \sum_{\type \in \mc{T}} x^{1,\type}_e - \frac{a_e\tilde{x}^2_e}{2} - b_e \sum_{\type \in \mc{T}} \tilde{x}^\type_e\right) \leq 0,
	\end{align*}
 where $x^{1,\type}_e$ denotes the total flow on edge $e$ by users of type
 $\type$ in the solution $\vec{x}^1$. Let us define $\Delta x^\type_e =
 \tilde{x}^\type_e - x^{1,\type}_e$, and $\Delta x_e = \tilde{x}_e - x^1_e$. By
 summing the above inequalities, the $a_e$ terms cancel out giving us
 \begin{align*}
 &\textstyle\sum_{e \in \E} b_e \sum_{\type \in \T} (\frac{\Delta x^\type_e}{r }
 -\Delta x^\type_e) = \sum_{e \in \E} b_e \sum_{\type \in \T} (\frac{1}{r} -1) \Delta x^\type_e
\end{align*}
where the right-hand side is less than or equal to zero.
Using the fact that $\sum_{\type}\Delta x^\type_e = \Delta x_e$ for all edges,
we get that
 \begin{align}
 \label{eqn_equilibrium_conditions}
\textstyle (\frac{1}{r}-1)  \sum_{e \in \E}b_e\Delta x_e \leq 0.
 \end{align}
Hence,
\begin{enumerate}
\item When $r > 1$,   $(\frac{1}{r}-1) < 0$ so that $\sum_{e \in \E}b_e\Delta x_e \geq 0$.

\item When $ r < 1$, $(\frac{1}{r}-1) > 0$ so that $\sum_{e \in \E}b_e \Delta x_e \leq 0$.
\end{enumerate}
We finish the proof by considering two separate cases: (case 1) $1 \leq r \leq
2$ and (case 2) $r<1$.

Let us consider (case 1) where $1 \leq r \leq
2$.
Applying Lemma~\ref{lem_technical_socialcost} to the instance $\mc{G}$ with
$\vec{x} = \vec{x}^1$, we obtain that
\begin{equation}
\label{eqn_socialcost2}
\textstyle\text{C}(\vec{\tilde{x}}) - \text{C}(\vec{x}^1) \leq (1-\frac{2}{r}) \sum_{e \in \E}b_e \Delta x_e. \end{equation}
 We claim that when $r \in (1,2]$, the right-hand side of
     \eqref{eqn_socialcost2} is lesser than or equal to zero. This is not
     particularly hard to deduce owing to the fact that $(1-\frac{2}{r}) < 0$ in the
     given range and that $\sum_{e \in \E}b_e \Delta x_e \geq 0$ as deduced from
     \eqref{eqn_equilibrium_conditions}. Therefore, $\text{C}(\vec{\tilde{x}}) -
     \text{C}(\vec{x}^1) \leq 0$, which proves the claim that
     uncertainty with a limited amount of caution helps lower equilibrium
     costs.

     Now, let us consider (case 2) where $r<1$.
 Applying Lemma~\ref{lem_technical_socialcost} to the instance $\mc{G}^1$ with
 $\vec{x}=\tilde{\vec{x}}$, and using the fact that $\Delta x_e = \tilde{x}_e -
 x^1_e$, we have that
 \begin{equation}
 \label{eqn_socialcost3}
 \textstyle\text{C}(\vec{x}^1) - \text{C}(\vec{\tilde{x}})\leq (\frac{2}{r}-1) \sum_{e \in \E}b_e \Delta x_e. \end{equation}
 Once again when $r < 1$, we know that $\frac{2}{r}-1 > 0$ and from \eqref{eq:eqn_equilibrium_inequality}, we can deduce that $\sum_{e \in \E}b_e\Delta x_e \leq 0$ in the given range. 
\end{IEEEproof}

The following corollary identifies a specific level of uncertainty at which the equilibrium solution is actually optimal.
\begin{corollary}
	\label{corr_optatpointfive}
Given an instance $\mc{G}$ of the multi-commodity routing game, let
$\tilde{\vec{x}}$ denote its Nash equilibrium and $\vec{x}^\ast$ denote the
socially optimal flow. If $r_\type = 2$ for all $\type \in \T$, then
$\text{C}(\tilde{\vec{x}}) = \text{C}(\vec{x}^\ast)$---i.e.~the equilibrium is socially optimal.
\end{corollary}
\begin{IEEEproof}
Suppose $r_\theta=r=2$. Then, applying Lemma~\ref{lem_technical_socialcost}, we have that
$\text{C}(\tilde{\vec{x}}) - \text{C}(\vec{x}^\ast) \leq - (\frac{2}{r} - 1) \sum_{e \in
\E } b_e -\Delta x_e = 0$. \end{IEEEproof}
\subsection{Price of Anarchy Under Uncertainty}
In Theorem~\ref{thm_instancewise}, we showed that the equilibrium cost under
uncertainty decreases (resp.~increases) when users are mildly cautious
(resp. not cautious) and all user types have the same level of uncertainty. This
naturally raises the question of quantifying the improvement (or degradation) in
equilibrium quality and whether uncertainty helps when the uncertainty parameter can
differ between user types. In the following theorem, we address both of these
questions by providing price of anarchy bounds as a function of the maximum uncertainty in the system and $\gamma$, which is the ratio between the minimum and maximum uncertainty among user types.
\begin{theorem}{(Price of Anarchy)}
	\label{thm:poa}
	For any multi-commodity routing game $\mc{G}$, the ratio between the social cost of the Nash equilibrium to that of the socially optimal solution is at most
	\begin{equation}
	\label{eqn:poa1}
   PoA(\mc{G}) = 4/(4\gamma \rmax - \rmax^2),
	\end{equation}
	where $\rmax = \max_{\type \in \T} r_\type$, and $\gamma = \frac{\min_\type r_\type}{\max_\type
    r_\type}$ as long as $\rmax < 4\gamma$.
\end{theorem}
%

\begin{figure}[t]
	\centering
	\begin{center}
	\includegraphics[width=0.9\columnwidth]{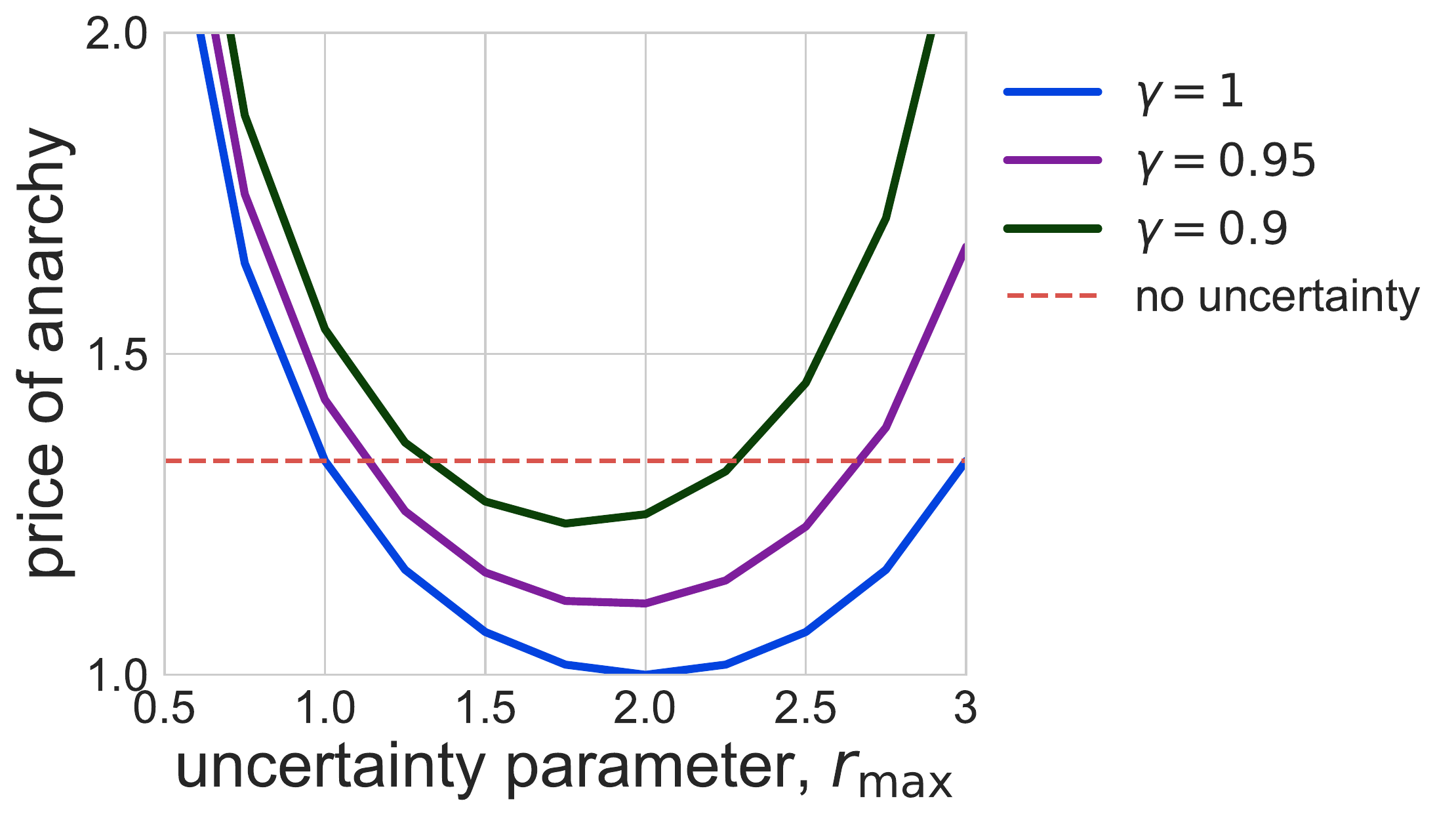}\end{center}
	\caption{\label{fig:poa}Price of anarchy as a function of $\rmax = \max_\type
		r_\type$ for multi-commodity selfish routing games for three different
		values of $\gamma = \frac{\min_\type r_\type}{\max_\type r_\type}$.  Smaller values of price of anarchy are favorable as they indicate that the
		cost of the equilibrium solution is comparable to the social optimum. In
		general, we observe that when $\rmax >1 $, the price of anarchy under
		uncertainty is smaller than that without uncertainty and vice-versa for $\rmax <
		1$. Yet, too much caution or large asymmetries in the uncertainty level
		across user types can also lead to poor equilibrium. For a fixed value of
		$\gamma$, we observe that the price of anarchy drops initially as $\rmax$ increases and then starts increasing once again beyond a certain threshold.}
\end{figure}

Before proving Theorem~\ref{thm:poa}, we remark on the price of anarchy and its
dependence on the level of uncertainty. The price of anarchy in \eqref{eqn:poa1} is plotted in Fig.~\ref{fig:poa} as a
function of $\rmax$ for three different values of $\gamma$. This result validates our message that uncertainty helps
equilibrium quality when users over-estimate their costs and hurts equilibrium
quality when users under-estimate their costs. To understand why, let us first consider the case of
$\gamma = 1$---i.e.~the uncertainty is the same across user types. We already
know that in the absence of uncertainty, the price of anarchy of multi-commodity
routing games with linear costs is given by $4/3$~\cite{roughgardenT02};
this can also be seen by substituting $\rmax = 1$, $\gamma = 1$ in
\eqref{eqn:poa1}. We observe that the price of anarchy is strictly smaller than
$4/3$ for $\rmax \leq 2$ and reaches the optimum value of $1$ at $\rmax = 2$ thereby confirming Corollary~\ref{corr_optatpointfive}. More interestingly, even when $\rmax > 2$, the price of anarchy with uncertainty is smaller than that without uncertainty, affirming our previous statement that cautious behavior helps lower congestion (at least in the worst case).

Similarly, as $\rmax$ decreases away from one, the price of anarchy increases nearly linearly.
In fact, our price of anarchy result goes one step beyond
Theorem~\ref{thm_instancewise} as it provides guarantees even when different
user types have different uncertainty levels. For example, when $\rmax = 1.5$,
and $\gamma = 0.9$---i.e.~$\min_\type r_\type ~\approx 1.35$---the price of
anarchy is $1.24$, which is still better than the price of anarchy without
uncertainty. 

The price of anarchy result reveals a surprising observation: as long as
$\rmax > 1$ and $\gamma$ is not too large, for any given instance $\mc{G}$ of the multi-commodity routing game, either the equilibrium quality is already good or uncertainty helps lower congestion by a significant amount. Therefore, uncertainty rarely hurts the quality of the equilibrium and often helps.

In order to prove Theorem~\ref{thm:poa}, we need the following technical lemma
whose proof is provided in  Appendix~\ref{app:proof:lem_technical_poa}.
  \begin{lemma}
  	\label{lem:lem_technical_poa}
  For any two non-negative vectors of equal length, $(x_1, x_2, \ldots, x_n)$ and
  $(x'_1, x'_2, \ldots, x'_n)$, let $x = \sum_{i=1}^n x_i$ and $x' = \sum_{i=1}^n
  x'_i$. Moreover, let $\vec{r} = (r_1, r_2, \ldots, r_n)$ be another vector of
  length $n$ whose entries are strictly positive. Then, if we let  $r_\ast =
  \min_i r_i$ and $r^\ast = \max_i r_i$, for any given function $f(y) = ay + b$ with $a,b \geq 0$, we have that
  \begin{equation}\textstyle \frac{f(x)x}{f(x')x' + \sum_{i=1}^n (x_i - x'_i) f(r_i x) } \leq
  4\left(4r_\ast - (r^\ast)^2\right)^{-1},
  \label{eqn_poaabstract}
  \end{equation}
  \end{lemma}
\begin{IEEEproof}[Proof of Theorem~\ref{thm:poa}]
    Consider some instance $\mc{G}$ of the
multi-commodity routing game with equilibrium $\tilde{\vec{x}}$. Then, adopting
the variational inequality conditions for a Nash equilibrium~\cite{thaiLB16,roughgardenT02}, for any other solution $\vec{x}$, and every user type $\type \in \T$, 
\begin{align}
\label{eqn_variational2}
\textstyle\sum_{e \in \E}\left(r_\type a_e \tilde{x}_e +
b_e\right)\left(\tilde{x}^{\type}_e - x^\type_e\right) &= \textstyle\sum_{e \in
\E}C_e(r_{\type}\tilde{x}_e)(\tilde{x}^{\type}_e - x^\type_e)\notag\\
&\leq 0.
\end{align}
must hold.
Fix any edge $e \in \E$.
Let $(x_1, \ldots, x_n) = (\tilde{x}^\type_e)_{\type \in \T}$,
$x=\sum_{\type \in \T}\tilde{x}_e^{\type}$, 
$(x'_1, \ldots, x'_n)
= (x^{*\type}_e)_{\type \in \T}$, and $x'=\sum_{\type \in \T}{x}_e^{\ast\type}$.
Applying Lemma~\ref{lem:lem_technical_poa} with $f(y) = C_e(y)$ and $\vec{r} =
(r_\type)_{\type \in \T}$, we have that 
\begin{align*}
\textstyle C_e(\tilde{x}_e)\tilde{x}_e \leq \zeta(C_e(x^\ast_e)x^\ast_e + \sum_{\type
\in \T}(\tilde{x}^{\type}_e -
x^{\ast\type}_e)C_e\left(x_er_\type\right) )
\end{align*}
where
\begin{equation}
    \textstyle\zeta=4/(4\rmin - \rmax^2) = 4/(4\gamma\rmax - \rmax^2),
    \label{eq:zeta}
\end{equation}
and where $\rmin = \min_{\theta \in \mc{T}}r_{\type}$ and $\rmax=\max_{\theta\in \T}r_\theta$. 

We claim that
\begin{align}
\textstyle\sum_{e \in \E}C_e(\tilde{x}_e)\tilde{x}_e &\textstyle \leq
\zeta\big(\sum_{e \in \E}\sum_{\type \in \T}(\tilde{x}^{\type}_e -x^{\ast\type}_e)\left(r_	\type a_e\tilde{x}_e + b_e\right) 
\nonumber \\
& \textstyle\qquad +\sum_{e \in
\E}C_e(x^\ast_e)x^\ast_e\big) \nonumber \\
& \textstyle\leq  \zeta \sum_{e \in \E}C_e(x^\ast_e)x^\ast_e. \label{eqn_poafinal}
\end{align}
Recall from \eqref{eqn_variational2} that for any type $\type \in
\T$, $\sum_{e \in \E}\left(r_\type a_e \tilde{x}_e +
b_e\right)\left(\tilde{x}^{\type}_e - x^\type_e\right)  \leq 0.$ Since the price of anarchy is defined as $\frac{\sum_{e \in \E}C_e(\tilde{x}_e)\tilde{x}_e }{\sum_{e \in E} C_e(x^\ast_e)x^\ast_e}$, a worst case bound of $\zeta$ follows from Equation~\eqref{eqn_poafinal}, giving us the theorem statement.
\end{IEEEproof}

It is worth mentioning that the proof of the above theorem carries over even
when each user type has a different uncertainty parameter on each edge. For
example, suppose that $r_{\type}(e)$ denotes the uncertainty parameter
corresponding to users of type $\type$ on edge $e \in \E$. In this general
model, a similar price of anarchy bound can be shown. Indeed, the price of
anarchy in this case is
\[\text{PoA}(\mc{G})=\textstyle\max_{e \in \E}4/(4\gamma(e)\rmax(e) - \rmax^2(e)),  \]
where $\gamma(e) = \min_{\type \in \mc{T}}r_{\type}(e)/(\max_{\type \in
    \mc{T}}r_{\type}(e))$ and $\rmax(e) = \max_{\type \in \mc{T}}r_{\type}(e)$.



\subsection{Extensions to Polynomial Cost Functions}
\label{sec:polynomial}
We now generalize our results in another direction by considering multi-commodity routing games where the edges have polynomial cost functions. Specifically, we consider games where the cost function on any edge $e \in \E$ is of the form $C_e(x) = a_ex^d + b_e$, such that  the degree $d \geq 1$, and $a,b \geq 0$. This class of polynomial cost functions is referred to as \emph{shifted monomials of degree $d$}\cite{colini2017demand}, and has received considerable attention in the literature owing to applications in transportation~\cite{manual1985special, altmanBJS02,calderone:2016aa}--- e.g., link performance as a function of the number of cars on a road can be modeled as a shifted monomial function of degree four~\cite{manual1985special}.

We now show that both of our results from this section extend gracefully to the case where the edges have shifted monomial cost functions. In fact, when the edge cost functions are super-linear, uncertainty leads to more favorable results than the linear cost case. Informally, as the degree of the monomial $d$ grows, we show that uncertainty leads to a decrease in the social cost for a larger range of the parameter $r_{\type}$. For convenience, we will assume that all edges have shifted monomial cost functions with the same degree $d \geq 1$.

We begin by a identifying the potential function that the Nash equilibrium solution minimizes for a generalization of the multi-commodity routing game where every edge $e \in \E$ has the true cost function $C_e(x) = a_e x^d + b_e$, which users of type $\type \in \mc{T}$ perceive as $\hat{C}^{\type}_e(x) = r_{\type} a_e x^d + b_e$.  Therefore, equilibrium existence is guaranteed for this more general class of functions.
\begin{equation}
\textstyle\Phi_{d,r}(\vec{x}) = \sum_{e \in \E}  \left(\frac{1}{d+1}a_ex^{d+1}_e + b_e
\sum_{\type \in \mc{T}} \frac{ 1}{r_{\type}} x^\type_e\right) \label{eq:potentialfunction_poly}
\end{equation}

The following result generalizes Theorem~\ref{thm_instancewise} to games with shifted monomial cost functions. 

\begin{proposition}
	\label{prop_instancewise_poly}
	Consider any given instance $\mc{G}$ of the multi-commodity routing game with shifted monomial cost functions of degree $d$ having
	Nash equilibrium $\tilde{\vec{x}}$ and the corresponding game instance
	$\mc{G}^1$, having no uncertainty, with Nash equilibrium $\vec{x}^1$. Suppose $r_\type=r$ for all $\type \in \T$.
	Then, the following hold:
	\begin{enumerate}
		\item $\text{C}(\tilde{\vec{x}}) \leq \text{C}(\vec{x}^1)$ if $1 \leq r \leq d+1$.
		\item $\text{C}(\tilde{\vec{x}}) \geq \text{C}(\vec{x}^1)$ if $0 \leq r \leq 1$.
	\end{enumerate}
	%
\end{proposition}
In comparison to Theorem~\ref{thm_instancewise} where uncertainty results in a decrease in social cost for $r \in [1,2]$, we observe (surprisingly) that the range of the uncertainty parameter under which caution yields an improvement in social cost  increases linearly with $d$, i.e., uncertainty is more helpful as the cost function is more convex. On the other hand, the under-estimation of costs increases social cost regardless of the degree of the polynomial. The key idea involved in the proof of Proposition~\ref{prop_instancewise_poly} is a strict generalization of Lemma~\ref{lem_technical_socialcost} via a potential function argument, to obtain the following difference in costs:  $\text{C}(\tilde{\vec{x}}) -  \text{C}(\vec{x}) \leq -\sum_{\type \in \T} (\frac{d+1}{r_\type} - 1)\sum_{e \in \E}b_e(\tilde{x}^{\type}_e - x^\type_e)$. Recall that in the previous inequality $\vec{\tilde{x}}$ is the equilibrium solution for a given instance $\mc{G}$ and $\vec{x}$ is an arbitrary feasible flow for the same instance. The rest of the proof is analogous to that Theorem~\ref{thm_instancewise}. 

Next, we generalize the price of anarchy bounds from Theorem~\ref{thm:poa}.

\begin{figure}[t]
	\centering
	\begin{center}
		\includegraphics[width=0.9\columnwidth]{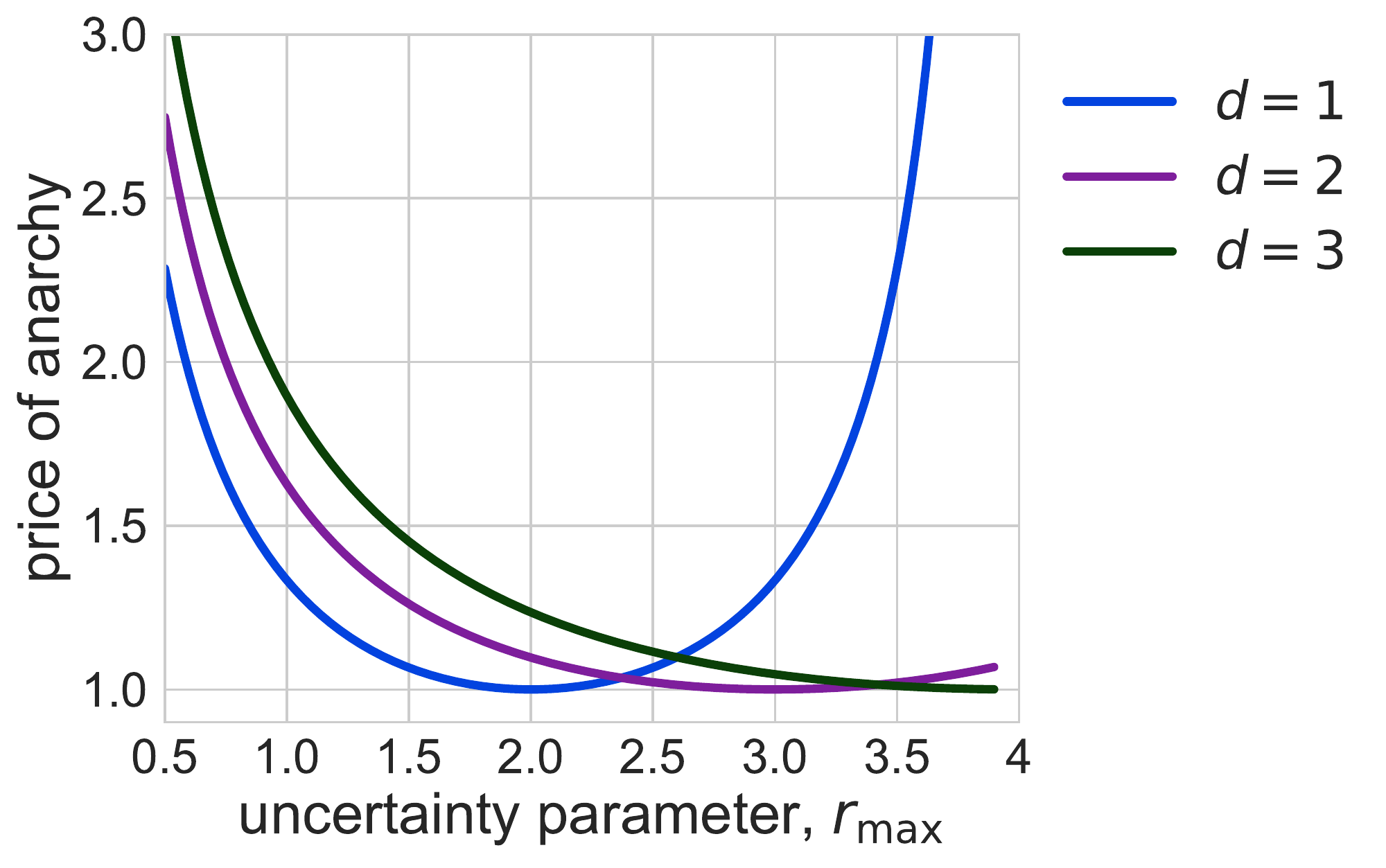}\end{center}
	\caption{\label{fig:poa_poly}Price of anarchy as a function of $\rmax = \max_\type
		r_\type$ for multi-commodity selfish routing games for three different
		values of $d$ when edge cost functions are shifted monomials of degree $d$.  Smaller values of price of anarchy are favorable as they indicate that the
		cost of the equilibrium solution is comparable to the social optimum. In
		general, we observe that when $\rmax >1 $, the price of anarchy under
		uncertainty is smaller than that without uncertainty and vice-versa for $\rmax <
		1$--- this is true for all values of $d$. Interestingly, for $d > 1$, uncertainty leads to near-optimal price of anarchy even for large values of $\rmax$ (e.g., $\rmax =3.5$)--- this is not the case when $d=1$. This validates our premise that as the degree of the polynomial cost function increases, user uncertainty becomes more helpful to the system welfare. }
\end{figure}

\begin{proposition}
	\label{poly:poa_poly}
	For any multi-commodity routing game $\mc{G}$ with shifted monomial cost functions of degree $d$, the ratio between the social cost of the Nash equilibrium to that of the socially optimal solution is at most
	\begin{equation}
	\label{eqn:poa1_poly}
	\displaystyle
	PoA(\mc{G}) = \frac{(d+1)^{(d+1)/d}}{\gamma \rmax(d+1)^{(d+1)/d}  - d\rmax^{(d+1)/d}},
	\end{equation}
	where $\rmax = \max_{\type \in \T} r_\type$, and $\gamma = \frac{\min_\type r_\type}{\max_\type
		r_\type}$ as long as $\rmax < (\frac{\gamma}{d})^d (d+1)^{(d+1)}$.
\end{proposition}
Substituting $d = 1$ in the above equation, we obtain Theorem~\ref{thm:poa} as a special case. In the absence of uncertainty, it is known that as $d$ grows large, the price of anarchy bound grows as $O(\frac{d}{\log(d)})$~\cite{roughgarden2007routing}. Assuming $\gamma = 1$, for a constant $\rmax$, the price of anarchy bound from Equation~\eqref{eqn:poa1_poly} grows as $O(\frac{d}{\rmax \log(d)})$. For a finite value of $d$, the improvement in the price of anarchy is much more significant as illustrated in Fig.~\ref{fig:poa_poly}.

\section{The Effect of Heterogeneity on Congestion}
\label{sec:heterogeneity}
Now that we have a better understanding of how uncertainty affects the
performance of the entire system as measured by the social cost, we move on to a
more nuanced setting where different users have different levels of uncertainty.
Our goal in this section is to understand the effect of this \emph{heterogeneity
in uncertainty} on the equilibrium congestion by studying the following two
questions: (i) how do the routing choices adopted by the uncertain users impact
the cost of the users without uncertainty, and (ii) when is the equilibrium
social cost of a system with heterogeneous uncertainties smaller than the social cost incurred when all users have no uncertainty? \subsection{Notation and Graph Topologies}
To isolate the effect of heterogeneity on equilibrium congestion, we restrict
our attention to a selfish routing game on an undirected network. There are two user types,
i.e.~$\mathcal{T} = \{\theta_1, \theta_2\}$. Users belonging to both these types seek to route their flow between source node $s$ and destination node $t$. Finally, the uncertainty levels for the two user types are specified as $r_{\type_1} = 1$ and $r_{\type_2} = r >0$, and so, users of type $\theta_1$ are without uncertainty. We refer to this as the \emph{two-commodity game with and without uncertainty}. We slightly abuse notation and use  $\mc{G}=\{G,\mathcal{T}, (s,t), (\mu_{\type_1}, \mu_{\type_2}), (1,r), (C_e)_{e\in \E}\}$ to refer to an instance of this game, and $\mc{P}$ to represent the set of $s$-$t$ paths in $G$.

Unlike the previous sections, where we made no assumptions on the graph structure, our next characterizations will depend crucially on the network topology. Specifically, we will consider two well-studied topologies: \emph{series-parallel} and \emph{linearly independent} graphs. We provide the respective definitions below.
\begin{defn}{(\emph{Series-Parallel}~\cite{milchtaich06})}
	\label{def:sp}
	An undirected graph $G$ with a single source $s$ and destination $t$ is said to be a series-parallel graph if no two $s$-$t$ paths pass through an edge in opposite directions.
\end{defn}
There are a number of other equivalent definitions for this class of graphs;
e.g., a graph is said to be series-parallel if it does not contain an embedded
\emph{Wheatstone network}~\cite{milchtaich06}. Series-parallel graphs are an
extremely well-studied topology that naturally arise in a number of applications
pertaining to network routing. We refer the reader to~\cite{acemogluO07,milchtaich06} for more details. 

\begin{defn}{(\emph{Linearly Independent}~\cite{milchtaich06})}
	An undirected graph $G$ with a single source $s$ and destination $t$ is said to be linearly independent if every $s$-$t$ path contains at least one edge that does not belong to any other $s$-$t$ path.
	\label{def:li}
\end{defn}
Our final definition involves a simple extension of the above topology to include linearly independent graphs connected serially. Formally, a graph $G = (V,\E)$ is said to consist of two sub-graphs $G_1 = (V_1, \E_1)$ with source-destination pair $(s_1, t_1)$ and $G_2 = (V_2, \E_2)$ with source-destination pair $(s_2, t_2)$ connected in serial if $V = V_1 \cup V_1$ with $t_1 = s_2$ and $\E = \E_1 \cup \E_2$.

\begin{defn}{(\emph{Serially Linearly Independent (SLI)})}
	An undirected graph $G$ with a single source $s$ and destination $t$ is said
    to belong to the serially linearly independent class if (i) $G$ is linearly independent or (ii) $G$ consists of two linearly independent graphs connected in serial.
	\label{defn_sli}
\end{defn}
This extended topology was first introduced in~\cite{acemoglu2016informational}. These three topologies are related as:
\begin{enumerate}
\item Every linearly independent graph belongs to the serially linearly independent class.
\item Every serially linearly independent graph belongs to the series-parallel class.
\end{enumerate}
The first part follows from Definition~\ref{defn_sli}. The second part is due to the fact that any linearly independent network is series-parallel (proved in~\cite{milchtaich06}) and the property that when two series-parallel graphs are connected in series, the resulting graph is also series-parallel.
Since series-parallel graphs happen to be the most general topology studied in this section, (without loss of generality) we focus on undirected graphs because every edge can be uniquely traversed only in one direction. %

\subsection{Impact of Uncertain Users on Users without Uncertainty}
We begin by studying what happens to the congestion cost faced by the users without uncertainty as the uncertainty level increases for the other users. This question is of considerable interest in a number of settings. For example, in urban transportation networks, the uncertainty about where a driver can find available street parking
can often cascade into increased congestion for other drivers leading to a
detrimental effect on the overall congestion cost~\cite{calderone:2016aa,dowling2017much,shoup2007gone,ratliffDMZ16}.



The following theorem shows a somewhat surprising result.
As long as the network topology is series-parallel, the aggregate cost felt by users of type $\type_1$ (users without any uncertainty) always reduces when users of $\type_2$ are uncertain about the costs. In other words, the behavior under uncertainty by one type of users always decreases the congestion costs of other types of users who do not face any uncertainty.

\begin{theorem}
	\label{thm:crossuncertainty}
Given an instance $\mc{G}$ of the two-commodity game with and without uncertainty such that the graph $G$ is series-parallel, let $\mc{G}^1$
denote a modified version of this instance with no uncertainty
(i.e.~$r_{\theta_1} = r_{\type_2} = 1$). Let
$\tilde{\vec{x}}$ and $\vec{x}^1$ denote the Nash equilibrium for the two
instances, respectively. Then,
\begin{equation}
\C^{\type_1}(\tilde{\vec{x}})  \leq \C^{\type_1}(\vec{x}^1).
\end{equation}
where $\text{C}^{\theta_1}(\vec{x})
= \sum_{e \in \E}C_e(x_e)x^{\theta_1}_e$ is the aggregate cost of users of type
$\type_1$.
\end{theorem}


\begin{IEEEproof} It is well-known that~\cite[Lemma
	3]{milchtaich06} for a series-parallel graph $G$ and
any two feasible flows $\vec{x}$,
$\vec{x}'$, there exists a $s$--$t$ path $p$ with $x_{p} > 0$, such that for
every edge $e \in p$, $x'_e \leq x_e$. Now, consider flows $\vec{x}^1$ and $\tilde{\vec{x}}$. Applying the previous property, we get that, there exists a path $p$ with $x^1_p > 0$ such that for all $e \in p$, $\tilde{x}_e \leq x^1_e$.

 We now bound both $\text{C}^{\type_1}(\tilde{\vec{x}})$ and
$\text{C}^{\type_1}(\vec{x}^1)$ in terms of the cost of the path $p$.
Specifically, note that in the solution $\vec{x^1}$, the path $p$ has non-zero
flow on it so that
\begin{equation}
\textstyle\text{C}^{\type_1}(\vec{x}^1) = \mu_{\type_1} \sum_{e \in p}C_e(x^1_e).
\end{equation}
However, in the solution $\tilde{\vec{x}}$, we know that every user of type
$\type_1$ is using a minimum cost path with respect to the true costs and therefore, the cost of any path used by type $\type_1$ is at least that of the path $p$. Formally,
\begin{equation}
\textstyle\text{C}^{\type_1}(\tilde{\vec{x}}) \leq \mu_{\type_1} \sum_{e \in
p}C_e(\tilde{x}_e) \leq \mu_{\type_1}\sum_{e \in p}C_e(x^1_e).
\end{equation}
The final inequality follows from the monotonicity of the cost functions and the
fact that $\tilde{x}_e \leq x^1_e$ for all $e\in p$. Therefore, we conclude that
$\text{C}^{\type_1}(\tilde{\vec{x}}) \leq \text{C}^{\type_1}(\vec{x}^1).$
\end{IEEEproof}

\subsection{Characterization of Instances where Heterogeneity Helps}
We now consider the impact of heterogeneity on the system performance as a whole and present a simple characterization based on the network topology and the level of uncertainty, where the presence of uncertainty (among a fraction of the user population)  results in a decrease in the equilibrium social cost. Specifically, we show that for SLI networks, as long as the uncertainty level of users belonging to type $\type_2$ is at most two (i.e., $1 \leq r = r_{\type_2} \leq 2$), the social cost of the equilibrium solution is always smaller than or equal to that of the equilibrium when there is no uncertainty. 

Before showing our theorem, we state the following technical lemma whose proof is deferred to Appendix~\ref{app:hetproof}.
\begin{lemma}
	\label{lem:flowrerouting}
	Given any instance $\mathcal{G}$ of the two-commodity routing game with and without uncertainty where the graph $G$ is linearly independent, let $\vec{y}$ and $\vec{y}^1$ denote the Nash equilibria of instances $\mathcal{G}$ and $\mathcal{G}^1$ respectively. Then, it must be the case that 	
	\begin{equation}
	y^{\type_1}_p \leq y^1_p \quad  \forall p \in \mathcal{P}.
	\end{equation}
\end{lemma}
Informally, the above lemma states that given equilibrium flows  $\vec{y}$ and $\vec{y}^1$ for any arbitrary instance $\mathcal{G}$ and its uncertainty-free variant $\mathcal{G}^1$, the equilibrium solutions must satisfy the property that for any path $p$, the flow on this path in the absence of uncertainty (instance $\mathcal{G}^1$) must be greater than or equal to its magnitude due to the uncertainty-free users in $\mathcal{G}$.

\begin{theorem}
	\label{thm_characterization}
	Consider any given instance $\mc{G}$ of the two-commodity game with and without uncertainty. Let  $\tilde{\vec{x}}$ denote the Nash equilibrium of this game and the corresponding game instance
	$\mc{G}^1$, having no uncertainty has Nash equilibrium $\vec{x}^1$. Then, as long as $G$ belongs to the serially linearly independent class and $1 \leq r\leq 2$,
	$$\text{C}(\tilde{\vec{x}}) \leq \text{C}(\vec{x}^1)$$
\end{theorem}
%
    \begin{IEEEproof}[Proof of Theorem~\ref{thm_characterization}]
    Each SLI network can be broken down into a sequence of linearly independent
    networks connected in series. Applying Definition~\ref{defn_sli}
    recursively, we get a sequence of linearly independent sub-graphs $G(1) =
    (V(1),\E(1)), G(2) = (V(2),\E(2)), \ldots, G(\ell) = (V(\ell), \E(\ell))$ with
    source-destination pairs $(t_0,t_1), (t_1, t_2), \ldots, (t_{\ell-1},
    t_\ell)$ respectively ( note that $t_0 = s, t_\ell = t$), that are connected
    in series---i.e., $G_1$ is connected in series with $G_2$ such that the destination $t_1$ for $G_1$ acts as the origin for $G_2$. By definition, the set of edges in these subgraphs are mutually disjoint.

	Secondly, given the equilibrium flow $\vec{\tilde{x}}$ on $G$ for instance
    $\mathcal{G}$,  we can divide this flow into components
    $(\tilde{\vec{x}}(1),\tilde{\vec{x}}(2), \ldots, \tilde{\vec{x}}(\ell))$
    such that for every $1 \leq i \leq \ell$, $\tilde{\vec{x}}(i)$ is the
    sub-flow of $\vec{\tilde{x}}$ on the graph $G(i)$, and for every $e \in \E(i)$, $\tilde{x}(i)_e = \tilde{x}_e$. Finally, it is not hard to see that $\vec{\tilde{x}}(i)$ must be an equilibrium of the sub-instance of $\mathcal{G}$ restricted to the graph $G(i)$.

%
    Given this \emph{decomposition}, we apply Lemma~\ref{lem:flowrerouting}
    to each $G(i)$. Consider any index $i$: since the graph $G(i)$ is linearly
    independent, we can apply Lemma~\ref{lem:flowrerouting} and get that for any
    $t_{i-1}$-$t_{i}$ path $p$ in $G(i)$,
$\tilde{x}(i)^{\type_1}_p \leq x^1(i)_p$.	

	Suppose that $\mathcal{P}(i)$ denotes the set of $t_{i-1}$-$t_{i}$ paths in $G(i)$. 	
	Then, by  Lemma~\ref{lem:techconvex} provided in
    Appendix~\ref{app:techclaims},
    \begin{align}
\textstyle	C(\vec{\tilde{x}}) & = \textstyle\sum_{e \in \E}C_e(\tilde{x}_e)\tilde{x}_e = \sum_{e
    \in \E}(a_e \tilde{x}_e + b_e)\tilde{x}_e \nonumber \\
		& \textstyle\leq \sum_{e \in \E}\left(C_e(x^1_e)x^1_e + (2a_e
        \tilde{x}_e + b_e)(\tilde{x}_e - x^1_e) \right)\notag\\
       & \leq \textstyle C(\vec{x}^1) + \sum_{e \in \E}(2a_e \tilde{x}_e + b_e)(\tilde{x}_e - x^1_e) \label{eqn_decompose0}
    \end{align}
   so that, using the above decomposition,
   \begin{align}
  C(\vec{\tilde{x}})  
	& \textstyle\leq C(\vec{x}^1) + \sum_{i=1}^{\ell}\sum_{e \in \E(i)}(2a_e
    \tilde{x}(i)_e + b_e)\nonumber\\
    &\qquad \cdot(\tilde{x}(i)_e - x^1(i)_e) \nonumber\\
	& =\textstyle C(\vec{x}^1)  + \sum_{i=1}^{\ell}\sum_{p \in
        \mathcal{P}(i)}\big(\sum_{e \in p}(2a_e \tilde{x}(i)_e + b_e)
        \big)\notag\\
        &\qquad\cdot(\tilde{x}(i)_p - x^1(i)_p) \nonumber
\end{align}
Therefore, to complete the proof it is sufficient to show that  $\sum_{p \in \mathcal{P}(i)}\left(\sum_{e \in p}(2a_e \tilde{x}(i)_e + b_e) \right)(\tilde{x}(i)_p - x^1(i)_p) \leq 0$ for all $1 \leq i \leq \ell$. 

Fix an arbitrary index $i$ and consider the corresponding graph $G(i)$ and flows $\vec{\tilde{x}}(i)$ and $\vec{x}^1(i)$.
Recall that $\vec{x}^1(i)$ is the equilibrium solution in the absence of uncertainty and therefore, minimizes the following potential function
	\begin{equation*}
\textstyle	\Phi_\vec{1}(\vec{x}) = \sum_{e \in \E(i)}(\frac{1}{2}a_e(x_e)^2 + b_e x_e).
\end{equation*}
so that, again by Lemma~\ref{lem:techconvex},
\begin{align}
	0 & \leq \Phi_\vec{1}(\vec{\tilde{x}}(i)) - \Phi_\vec{1}(\vec{x}^1(i)) \nonumber \\
	& =\textstyle \sum_{e \in \E(i)}\big(\frac{a_e}{2}(\tilde{x}(i)_e)^2 + b_e
    \tilde{x}(i)_e \notag\\
    &\qquad\textstyle- \frac{a_e}{2} (x^1(i)_e)^2 - b_e x^1(i)_e\big) \nonumber\\
	& \leq\textstyle \sum_{e \in \E(i)} (a_e \tilde{x}(i)_e + b_e)(\tilde{x}(i)_e - x^1(i)_e) 
    \label{eqn_potential1}.
	\end{align}
Next, we claim that the term $\sum_{e \in \E(i)}(ra_e \tilde{x}(i)_e +
b_e)(\tilde{x}(i)_e - x^1(i)_e) \leq 0$. Indeed,  for convenience define
$D_r=\sum_{e \in \E(i)}(ra_e \tilde{x}(i)_e + b_e)(\tilde{x}(i)_e - x^1(i)_e)$
so that
	\begin{align}
	D_r & = \textstyle\sum_{p \in \mathcal{P}(i)} \sum_{e \in p}(ra_e \tilde{x}(i)_e +
    b_e)(\tilde{x}(i)_p - x^1(i)_p)\notag \\
	& = \textstyle\sum_{p \in \mathcal{P}(i)} \sum_{e \in p}(ra_e \tilde{x}(i)_e
    + b_e)(\Delta x_p) \notag\\
	& = \textstyle\sum_{p \in \mathcal{P}(i)^-}\sum_{e \in p}(ra_e \tilde{x}(i)_e + b_e)(\Delta x_p) \nonumber \\ 
    &\quad +\textstyle  \sum_{p \in \mathcal{P}(i)^+}\sum_{e \in p}(ra_e \tilde{x}(i)_e + b_e)(\Delta x_p)\label{eqn_potentialr},
	\end{align}
	where $\Delta x_p = \tilde{x}(i)_p - x^1(i)_p$ and $\mathcal{P}(i)^{-},
    \mathcal{P}(i)^+$ refer to the set of the paths where $\Delta x_p \leq 0$
    and $\Delta x_p > 0$, respectively. From Lemma~\ref{lem:flowrerouting}, we know that $\tilde{x}(i)^{\type_1}_p \leq x^1(i)_p$ for all $p \in \mathcal{P}(i)$. Therefore, if $\Delta x_p = \tilde{x}(i)^{\type_1}_p + \tilde{x}(i)^{\type_2}_p - x^1(i)_p > 0$ for any $p \in \mathcal{P}(i)$, it must be the case that $\tilde{x}(i)^{\type_2}_p > 0$ for that path. In other words, we have that for every $p \in \mathcal{P}(i)^+$, $\tilde{x}(i)^{\type_2}_p > 0$. Recall that $\tilde{x}(i)^{\type_2}_p$ denotes the total flow due to the users with uncertainty on path $p$.

	We make the following two observations that help us simplify \eqref{eqn_potentialr}.
            First, by flow conservation, we know that $\sum_{p \in \mathcal{P}(i)}\Delta x_p = 0$. 
		Second, for every $p \in \mathcal{P}(i)^+$, we have that $\tilde{x}(i)^{\type_2}_p > 0$, and so $p$ is a min-cost path for the users with uncertainty in the flow $\vec{\tilde{x}}(i)$ . This in turn implies that for any $p' \in \mathcal{P}(i)$ and $p \in \mathcal{P}(i)^+$, the following inequality is valid:
		\begin{equation}
	\textstyle	\sum_{e \in p}(ra_e \tilde{x}(i)_e + b_e) \leq \sum_{e \in p'}(ra_e \tilde{x}(i)_e + b_e).
		\end{equation}
	Define $c^r_p := \min_{p \in \mathcal{P}(i)}\sum_{e \in p}(ra_e
    \tilde{x}(i)_e + b_e)$. Then, using the fact $\Delta x_p \leq 0$ for all $p
    \in \mathcal{P}(i)^-$ and that $\sum_{e \in p}(ra_e \tilde{x}(i)_e + b_e)
    \geq c^r_p$ for all $p \in \mathcal{P}(i)^-$,
    \eqref{eqn_potentialr} can be rewritten as
	\begin{align}
	D_r & = 
	\textstyle \sum_{p \in \mathcal{P}(i)^-}\sum_{e \in p}(ra_e \tilde{x}(i)_e + b_e)(\Delta x_p)\nonumber \\
    & \quad+ \textstyle\sum_{p \in \mathcal{P}(i)^+}\sum_{e \in p}c^r_p(\Delta
    x_p)\notag\\
	& \leq \textstyle\sum_{p \in \mathcal{P}(i)^-}\sum_{e \in p}c^r_p(\Delta
    x_p) + \sum_{p \in \mathcal{P}(i)^+}\sum_{e \in p}c^r_p(\Delta x_p)\notag\\
	& \textstyle= c^r_p \sum_{p \in \mathcal{P}(i)}\Delta x_p = 0\label{eqn_rinequality}
	\end{align}
    so that
    \begin{equation}
    \textstyle    \sum_{e \in \E(i)}(ra_e \tilde{x}(i)_e + b_e)(\tilde{x}(i)_e - x^1(i)_e) \leq 0
        \label{eq:rinequality}
    \end{equation}

    Summing inequalities \eqref{eq:rinequality} and \eqref{eqn_potential1},  we
    see that the $b_e$ terms cancel out, leaving us with
    \begin{equation*}
   \textstyle     (r-1)\sum_{e \in E(i)}a_e\tilde{x}(i)_e(\tilde{x}(i)_e - x^1(i)_e)  \leq 0
    \end{equation*}
    which implies that 
    \begin{equation}
        \textstyle\sum_{e \in \E(i)}a_e\tilde{x}(i)_e(\tilde{x}(i)_e - x^1(i)_e) \leq 0
        \label{eq:a}
    \end{equation}
since $r>1$. This, in turn, implies that
\begin{equation}
 \textstyle   (2-r) \sum_{e \in \E(i)}a_e\tilde{x}(i)_e(\tilde{x}(i)_e - x^1(i)_e)  \leq 0
    \label{eq:finalrinequality}
\end{equation}
since $r\leq 2$.
Adding
    \eqref{eq:finalrinequality} to \eqref{eq:rinequality}, we get that
	\[\textstyle \sum_{e \in E(i)}(2a_e \tilde{x}(i)_e + b_e)(\tilde{x}(i)_e -
    x^1(i)_e) \leq 0.\]
	The above inequality in combination with \eqref{eqn_decompose0} yields the theorem.\end{IEEEproof}
\subsection{Tightness of Results: Negative Examples}
Our central result in this section (Theorem~\ref{thm_characterization}) shows
that for networks having the SLI topology with a limited amount of uncertainty
$(r \in [1,2])$, the social cost at equilibrium in the presence of uncertainty
is always smaller than or equal to the social cost without any uncertainty.
Naturally, this raises the question of whether the result is tight---i.e.~what happens when the graph does not belong to the SLI class or if $r > 2$. In this section, we show the tightness of our results by illustrating examples where uncertainty leads to an increase in congestion costs when either one of the above requirements fail.

\subsubsection{Series-Parallel Networks where Uncertainty Hurts}
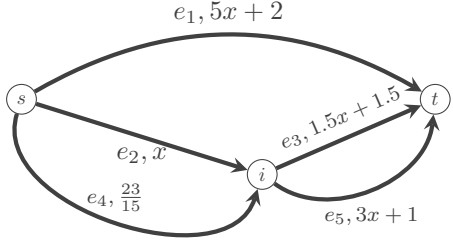
\begin{figure}[t!]
	\centering
	\begin{tikzpicture}
	\begin{scope}

	\node (fr) at (0, -0.4) {};

	\begin{scope}
	\node (arbit1) at (1.5, 0.5) {};
	\node (arbit2) at (1.5, -1.8) {};
	\draw (fr) +(-1, -1) node[labeledNodeS] (org){$s$};
	\draw (fr) +(4.5 ,-1) node[labeledNodeS] (v1) {$t$}
	edge[normalEdgeF, <-,bend right] node[above] {$e_1, 5x + 2$} (org) ;



	\draw (fr) +(2.2,-2)  node[labeledNodeS] (dest){$i$}
	edge[normalEdgeF, ->] node[above, sloped] {\footnotesize $e_3, 1.5x + 1.5$} (v1) {}
	edge[normalEdgeF, <-] node[below] {$e_2, x$} (org)
	edge[normalEdgeF, <-, bend left=90] node[above] {\footnotesize $e_4, \frac{23}{15}$} (org)
		edge[normalEdgeF, ->, bend right = 60] node[below=0.1] {\footnotesize $e_5, 3x+1$} (v1);



	\end{scope}

	\end{scope}
	\end{tikzpicture}
	\caption{\label{fig:sp_baduncertainty} An example of a series-parallel
network that is not SLI, where all traffic originates at node $s$ and terminates
at $t$. The label on each edge represents its identity and cost. For the
two-commodity game with and without uncertainty, the total population is $1$ and
a population of $\epsilon = 0.05$ users have an uncertainty factor of $r=2$.  We
label the five source-destination paths as $p_1$ ($e_1$), $p_2 (e_2,e_3), p_3
(e_2, e_5), p_4(e_4,e_3)$, and $p_5 (e_4, e_5) $. The social cost for this
example in the presence of uncertainty is strictly larger than the social cost
in the absence of uncertainty, indicating that for networks that violate the SLI
condition, uncertainty does not always improve equilibrium quality. }  		 		
\end{figure}

Fig.~\ref{fig:sp_baduncertainty} depicts an example of a network that is series-parallel but not SLI, where the equilibrium cost when a small fraction of users have an uncertainty of $r=2$ becomes strictly larger than the cost when all users have no uncertainty. The details of the equilibrium solutions for this example are listed in Table~\ref{table_ex:sli}.  Contrasting our result in Theorem~\ref{thm_characterization}, this example indicates that for networks that violate the SLI topology, the helpful effects of uncertainty are not guaranteed even when $r \in [1,2]$. Note that when $r \in [1,2]$, uncertainty can lead to an increase in social cost only when it is exhibited by a (small) fraction of users. This is because when the entire population has an uncertainty parameter of $1 \leq r \leq 2$, we know by Theorem~\ref{thm_instancewise} that the social cost must decrease for all graph topologies.

\begin{table}[h]
	\renewcommand{\arraystretch}{1.5}
	\centering
	\caption{	\label{table_ex:sli}A detailed description of  the flow on each path and social cost in the absence and presence of uncertainty: in the latter case, a small population of $\epsilon = 0.05$ users have an uncertainty of $r = 2$. The equilibrium flows for the paths that are not specified are zero. Note that $\frac{591}{200} = 2.955 > \frac{62}{21} \approx 2.9524$ and therefore, the equilibrium social cost in the absence of uncertainty is strictly smaller.}
    \begin{tabular}{|l||l|r|}
		\hline
		\textbf{Instance} & \textbf{Equilibrium Flow} & \textbf{Cost} \\
		\hline\hline
        Certain & $x_{p_1} = \frac{4}{21}$, $x_{p_2} = \frac{3}{7}$, $x_{p_3} =
        \frac{8}{21}$ &
    $62/21$\\\hline
    Uncertain   & $x_{p_1} = \frac{11}{60}$, $x_{p_2} = \frac{23}{60}$, $x_{p_3}
    = \frac{23}{60}$,
    $x_{p_4}=\frac{1}{2}$ & $591/200$ \\
\hline
	\end{tabular}

\end{table}


\subsubsection{Networks with Large Uncertainty, i.e., $r > 2$}
We now provide an example of a simple two-link Pigou network where the equilibrium social cost with uncertainty $r = 3$ is strictly larger than the equilibrium cost without any uncertainty. This example illustrates that even for networks that fall within the SLI class (parallel links are the most trivial class of networks), uncertainty can be detrimental to social cost when $r > 2$.

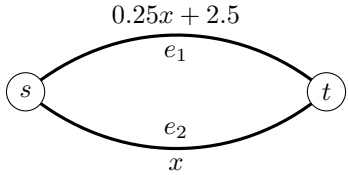
\begin{figure}[th!]
	\centering
	\begin{tikzpicture}
	\tikzset{edge/.style = {->,> = latex'}}
	\begin{scope}
	\coordinate (one)    at (2,0);
	\coordinate (oneb)   at (2,0.05);
	\coordinate (two)    at (0,0);
	\coordinate (three)  at (-2,0);
	\coordinate (threeb) at (-2,0.05);

	\path (three) edge[very thick, bend right=40,->] node[below] {$x$} node[above] {$e_2$} (one);
	\path (three) edge[very thick, bend left=40,->] node[above]{$0.25x + 2.5$} node[below] {$e_1$}(one);

	\filldraw[draw=black,fill=white] (one)   circle (0.1in);
	\filldraw[draw=black,fill=white] (three) circle (0.1in);
	\node at (one)   {$t$};
	\node at (three) {$s$};
	\end{scope}

	\end{tikzpicture}
	\caption{\label{fig:pigou_badexample}A two link Pigou network.
We consider a two-commodity game with and without uncertainty, where the total
population is one: a population of $0 < \epsilon \leq 1$ users have an
uncertainty parameter of $r = 3$. For every $\epsilon > 0$, the equilibrium social cost without uncertainty is strictly smaller than the social cost with uncertainty.}
\end{figure}

Consider the instance depicted in Fig.~\ref{fig:pigou_badexample}. In the
absence of uncertainty, the entire population uses $e_2$,
resulting in a social cost of $C_{e_2}(1) = 1$. Next, suppose that a population
of $\epsilon \leq 1$ users have an uncertainty level of $r = 3$. If $\epsilon \leq
2/15$, then all $\epsilon$ users route their flow on $e_1$, and
the social cost of the equilibrium solution with uncertainty is given by
$C_{e_1}(\epsilon)\epsilon + C_2(1-\epsilon)1-\epsilon = 1 + 0.5\epsilon +
1.25\epsilon^2$ which is strictly greater than $C_{e_2}(1)$, the social cost
without uncertainty. When $\epsilon > 2/15$,  only $2/15$--th of the users
prefer $e_1$, and the increase in social cost follows from the $\epsilon = 2/15$ case.

\section{Case Studies}
\label{sec:simulations}
%

\begin{figure}[t]
	\centering
	\begin{tikzpicture}
	\begin{scope}

	\node (fr) at (0, -0.4) {};

	\begin{scope}
	\node (arbit1) at (1.5, 0.5) {};
	\node (arbit2) at (1.5, -1.5) {};
	\draw (fr) +(0, -0.0) node[normalNodeS] (s1) {};
	\draw (s1) +(1.5, 0.5) node[normalNodeS] (v1) {}
	edge[normalEdgeF, <-,green!50!black] node[above] {} (s1) ;

	\node[above right=0.0cm and 2.7cm of s1] {$o_1$};
	\draw (s1) +(1.5, -0.5) node[normalNodeS] (v2) {$b$}
	edge[normalEdgeF, <-,green!50!black] node[below] {} (s1) ;
	\draw (s1) +(3, 0) node[normalNodeS] (t1) {}
	edge[normalEdgeF, <-,green!50!black] node[above,green!50!black] {} (v1)
	edge[normalEdgeF, <-,green!50!black] node[below] {} (v2);

	\draw (s1) +(-1, -0.75) node[labeledNodeS] (org){$s$}
	edge[normalEdgeF, ->] node[below] {} (s1) ;

	\draw(s1) +(1.5,-1.5) node[normalNodeS] (mid){}
	edge[normalEdgeF, <-] node[above] {} (org);

	\draw (s1) +(4,-0.75)  node[labeledNodeS] (dest){$t$}
	edge[normalEdgeF, <-] node[below] {} (t1)
	edge[normalEdgeF, <-] node[above] {} (mid);

	\draw(s1) +(2.5,-2.0) node[labeledNodeS] (pg) {$t_2$}
	edge[normalEdgeF, <-,bend left] node[below right=-0.0cm and 0.7cm] {Garage} (mid);
	\node[below right=0.85cm and 1.0cm of s1] {$o_2$};
	\node[green!50!black](arbit3) at (1.5, 0.5) {On-Street Parking};

	\end{scope}

	\end{scope}
	\end{tikzpicture}
	\caption{\label{fig:parkingrouting} A special case of our general
		multi-commodity network with two types of users: parking users and through
		traffic. All of the network traffic originates at the source node $s$. Users
		belonging to the through traffic population simply select a (minimum-cost) path from
		$s$ to $t$ and incur the latencies on each link. The parking users
		select between one of two parking structures: on-street parking (indicated
		in \textcolor{green!50!black}{green}) with
		additional circling costs and off-street parking (e.g., parking garage). }
\end{figure}
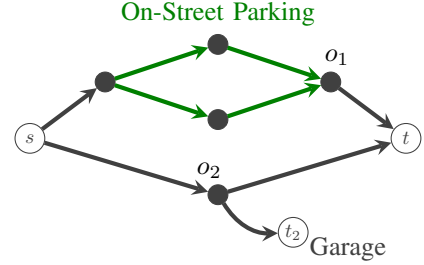

In this section, we present our main simulation results on both stylistic as
well as realistic urban network topologies comprising of two types of users---i.e.~\emph{through traffic, parking users} (types $\type_1,
\type_2$, respectively)---each
associated with a single commodity. We consider the more general edge-dependent uncertainty model that was briefly mentioned at the end of Section~\ref{sec:theory} and assume that the parking users have different uncertainty levels on different parts of the network and the through traffic does not suffer from uncertainty at all. For any given edge $e \in \E$, we assume that $C^{\type}_e(x)$ and $\hat{C}^{\type}_e(x)$  denote the actual and perceived cost of edge $e$ by users of type $\type$. We vary the level of uncertainty faced by the parking users, and observe its effect on the social cost at equilibrium.

Despite the generality of the model considered here---different user types have
different beliefs and the level of uncertainty depends on the edge under
consideration---our simulations validate the theoretical results presented in
the previous section. In particular, when the parking users are cautious (over-estimate the congestion cost due to cruising and waiting for a parking spot), we observe
that the social cost of the equilibrium decreases in both our experiments. On
the other hand, the behavior of parking users who under-estimate the cost
incurred due to searching for a parking spot leads to a worse-cost equilibrium.






\subsection{Effect of Uncertainty on On-Street vs Garage Parking}
\label{sec:simulations_simple}
Inspired by the work in~\cite{calderone:2016aa} which  provides a framework for
integrating parking into a classical routing game that abstracts route choices
in urban networks,
we begin with a somewhat stylized example of an urban network, depicted and
described in Fig.~\ref{fig:parkingrouting}. The users looking for a parking spot
are faced with two options: (i) \emph{on-street parking} which, as in reality,
is cheaper but leads to larger wait times due to cruising in search of parking; (ii) an \emph{off-street or a private garage option} that is much easier to access (in terms of wait times) at the expense of a higher price.

\begin{figure*}[t]
	\centering
	\includegraphics[width=0.7\textwidth]{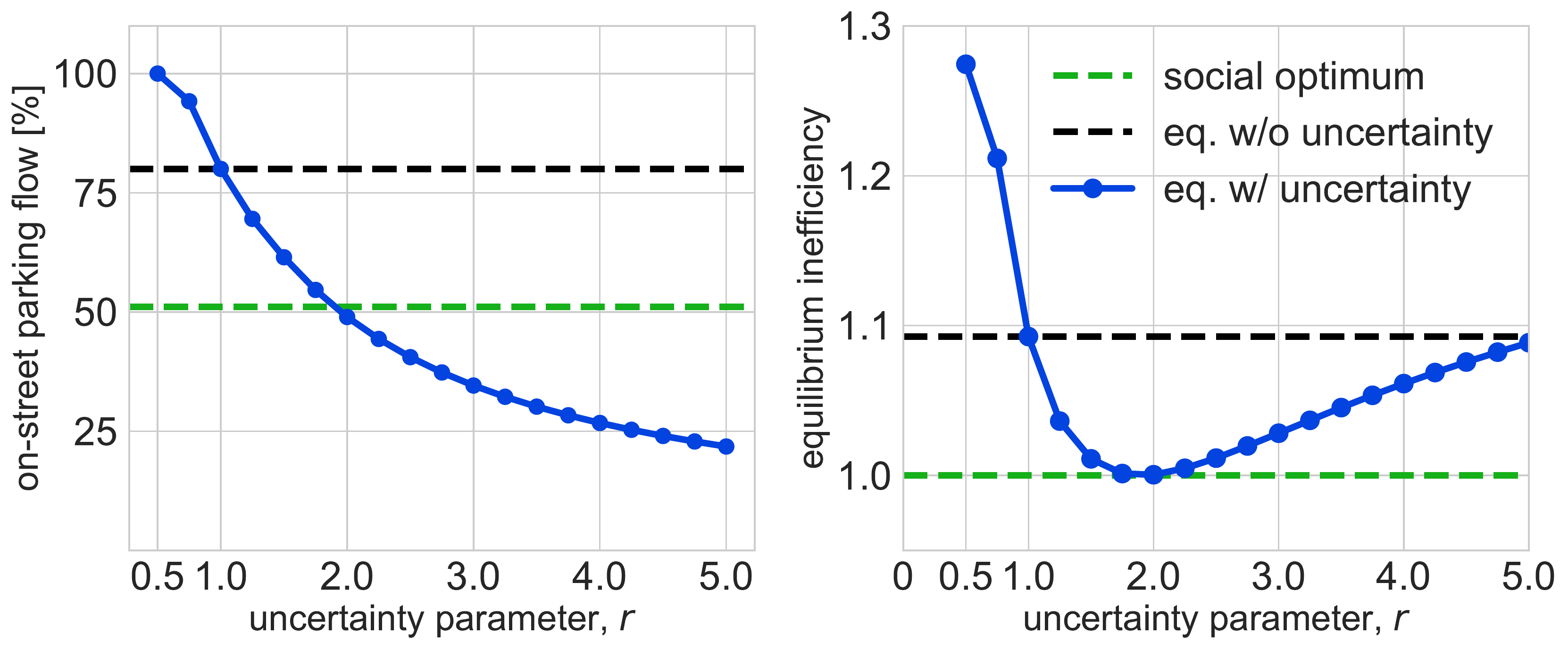}
	\caption{The left plot shows the on-street parking population flow (mass) under
		the social optimum and the Nash equilibrium with and without uncertainty as
		$r$, the level of uncertainty of the parking users, varies. The right plot shows the
		equilibrium quality as measured by the ratio of the social cost of the Nash equilibrium to that of the social optimum 
as a function of $r$.  True costs of off-street options are notoriously more
expensive than on-street options. As $r$ increases, we observe from the left plot that more users
move away from on-street parking, which in turn affects the social cost as seen
from the right plot.  In particular, when $r > 1$, the users over-estimate both
on-street and garage wait times leading to a decrease in social cost as more users
choose the garage option. On the other hand, for $r < 1$, users view the garage
option adversely, which leads to more congestion. We note that these
observations
are consistent with empirically observed human behavior~\cite{tversky:1991aa} suggesting that an
increase in cost from a  lower point of reference is seen as more of a loss than a
marginal increase in cost from a higher point of reference. At $r \approx
2$, the users divide themselves optimally between the on-street and garage
options leading to full efficiency; as $r$ increases beyond $2$, the
	equilibrium is still better than the case without uncertainty, yet no longer optimal.}
	\label{fig:parkingpop}
\end{figure*}

To understand the costs faced by the parking users (type $\theta_2$),  let $\mc{E}_{os}$ be the set of edges in the on-street parking structure (the \textcolor{green!50!black}{green} edges in Fig.~\ref{fig:parkingrouting}). For
parking users  that select the on-street option, the
cost on edges $e\in \E_{os}$ are of the form
\begin{equation}
C_e^{\theta_2}(x_e)=C_{e,\ell}^{\theta_2}(x_e)+C_{e,os}^{\theta_2}(x_e),
\label{eq:totalcoston}
\end{equation}
where $C_{e,\ell}^{\theta_2}(x_e)=a_ex_e+b_e$ is the travel latency part of the
cost and $C_{e,os}^{\theta_2}(x_e)=a_{os}x_e + b_{os}$ is the parking part of the
cost. Fig.~\ref{fig:parkingrouting} is easily converted into a two-commodity network by
creating a \emph{fake} edge $\tilde{e}$ from node $o_1$ to $t_2$ that has the accumulated
parking costs from the edges in $\E_{os}$---i.e.
\begin{equation}
C_{\tilde{e}}^{\theta_2}(x_{\tilde{e}})=\textstyle\sum_{e\in \E_{os}}
C_{e,os}^{\theta_2}(x_{\tilde{e}})=\bar{a}_{os}x_{\tilde{e}}+\bar{b}_{os}.
\label{eq:newcost}
\end{equation}
Then, the costs on edges in $\E_{os}$ are re-defined to only contain the travel
latency component of the cost, and this is the same for both types of users: for
$e\in \E_{os}$,
\begin{equation}
C_e^{\theta_1}(x_e)=C_e^{\theta_2}(x_e)=a_ex_e+b_e.
\label{eq:travellat}
\end{equation}
For the off-street parking structure, the edge, say $e'$, from $o_2$ to $t_2$ has cost
\begin{equation}
C_{e'}^{\theta_2}(x_{e'})=a_{pg} x_{e'} + b_{pg}
\label{eq:parkinggaragecost}
\end{equation}
and all other edges in the network have costs
\begin{equation}
C_e^{\theta_1}(x_e)=C_e^{\theta_2}(x_e)=a_ex_e+b_e.
\label{eq:otheredges}
\end{equation}

The price on-street is generally lower than that of a private garage, whereas
the inequality is reversed for wait times. Parking garages typically comprise of a larger capacity than on-street parking options and one expects this to be reflected in the $\bar{a}_{os}$ and $a_{pg}$ terms.

The uncertainty is faced by users of type $\type_2$ only on the costs pertaining to parking such that the congestion-dependent component of their parking cost is multiplied by a parameter $r > 0$. This captures the notion that the users may face uncertainty
regarding the number of other users competing for the same parking spots(s) or
the capacity of each structure. In the converted two commodity network, this translates to
\begin{equation*}
\hat{C}_{\tilde{e}}^{\theta_2}(x_{\tilde{e}})=r\bar{a}_{os}x_{\tilde{e}}+\bar{b}_{os}
\end{equation*}
where $\tilde{e}$ is the fake edge from $o_1$ to $t_2$ and
\begin{equation*}
\hat{C}_{e'}^{\theta_2}(x_{e'})=r a_{pg}x_{e'}+b_{pg}
\end{equation*}
where $e'$ is the edge from $o_2$ to $t_2$. As mentioned previously, the through
traffic population (type $\type_1$) does not suffer from uncertainty on any of its edges and therefore, $r_{\type_1}(e) = 1$ with respect to every edge $e$.






Note that although the uncertainty only applies to the parking
users, their decision also affects the through traffic users as both user types
share a subset of the edges on their routes. This phenomenon has been
commonly observed in cities where non-parking users often face congestion
due to the circling behavior of drivers looking for parking~\cite{shoup2007gone}.

For the simulations, we assume that there is a total population mass of $2$ originating at the source node
$s$, comprising of an equal number of parking  and through traffic users. The edge
congestion functions are selected uniformly at random from a suitable range.
The costs on the parking structures are set as follows:
\begin{enumerate} 
	\item  \emph{On-Street Parking}: The on-street parking structure is assumed
	to contain a fixed number of parking spots, and the parameter $\bar{a}_{os}$ is
	chosen to be inversely proportional to this quantity. The constant term
	$\bar{b}_{os}$ captures the price paid by drivers for on-street parking and is
	selected based on parking prices commonly used in large cities multiplied by a constant that captures how users tend to trade-off between time (congestion) and money.
	\item \emph{Off-street Parking}: Off-street parking is assumed to have a
	large number of available parking spots; thus, we set $a_{pg}  = 0$. The
	parameter $b_{pg}$ is set to be the price of off-street  parking (e.g., a garage) multiplied by the same trade-off parameter as above.
\end{enumerate}



%

Fig.~\ref{fig:parkingpop} shows how the parking users divide themselves among
the on-street and garage option (left plot) and how this affects equilibrium
quality as $r$ varies (right plot). From the left plot, we observe that at the
social optimum, approximately $51$\% of the parking population prefers on-street parking. With no uncertainty (i.e.~when $r=1$), at the Nash equilibrium more parking users (around $80$\%) gravitate towards the
cheaper on-street option leading to higher congestion and inefficiency---that is, even
without uncertainty, the system is inefficient as is expected. This is reminiscent of the classic Pigou
example~\cite{roughgarden2007routing} in traffic networks.

As $r$ increases---that is, as
users become more cautious in their beliefs about parking congestion---more users start
flocking to the off-street option due to the fact that they perceive a
multiplicative increase in the congestion-dependent term. This results in an
improvement to efficiency. On
the contrary, for users who tend to under-estimate parking costs $(r < 1)$, the appeal
of parking in off-street options decreases and more users flock towards
on-street parking leading to increased congestion and poor equilibrium quality.

The effect of the above behavior on equilibrium quality can be seen in the
right-hand plot in Fig.~\ref{fig:parkingpop}. The graph indicates that a cautious approach under uncertainty  (users over-estimating costs) helps improve equilibrium quality,
whereas  lack of caution  results in enhanced congestion in the network. In fact, we notice when $r < 1$, the price of anarchy increases in a rather steep fashion as $r$ decreases.  Finally, we observe that even though our theoretical results guarantee an improvement in equilibrium congestion only in the range of $1 \leq r \leq 2$, our simulation results lend credence to the claim that even highly cautious behavior ($r > 2$) can lead to a decrease in the social cost i.e., users over-estimating the costs in the face of uncertainty is always preferable to a lack of uncertainty. 
\subsection{Parking vs.~Through Traffic in Downtown Seattle}
\begin{figure}
	\centering
	
    \hfill\subfloat[Flow intensity for
	$r=1$]{\includegraphics[width=0.4\columnwidth]{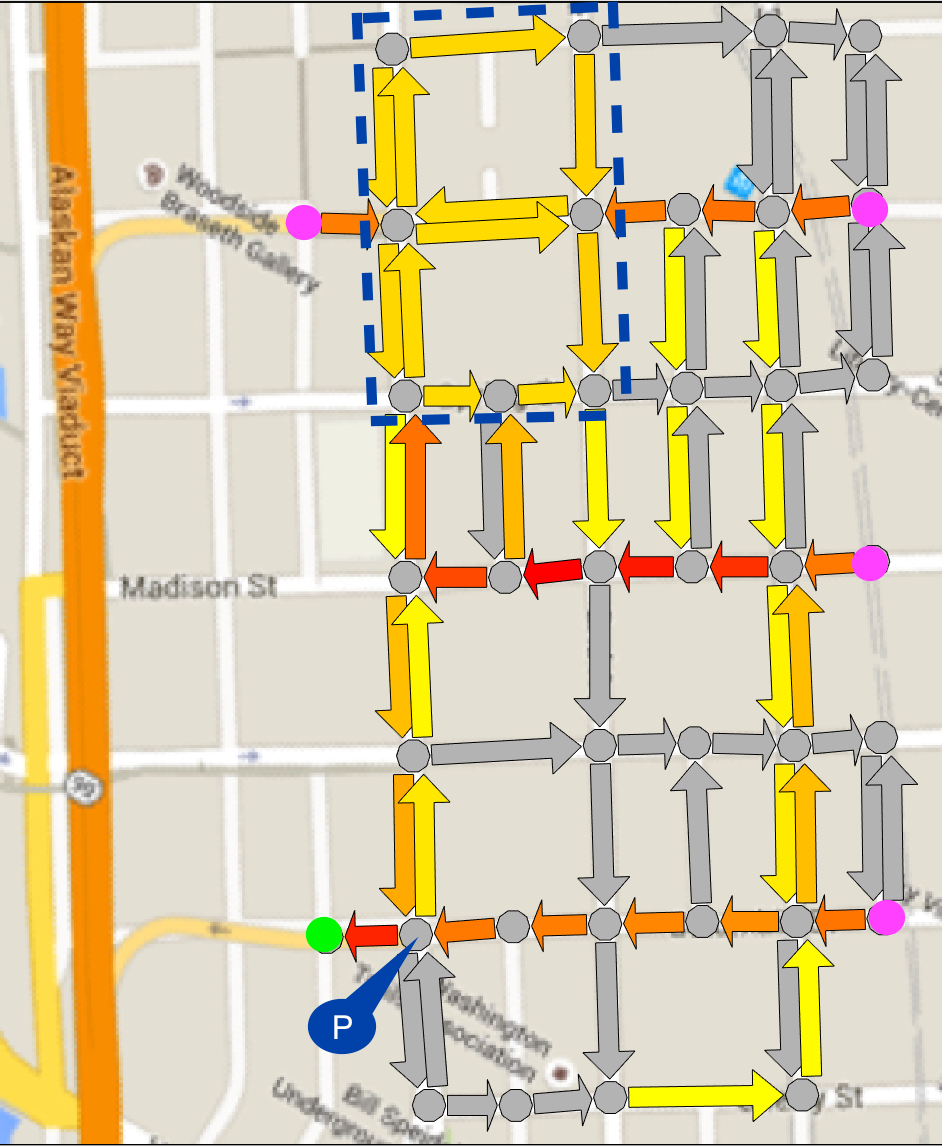}\label{fig:downtown_nouncertainty}
	}\hfill
	\subfloat[Flow intensity for
    $r=1.5$]{\includegraphics[width=0.4\columnwidth]{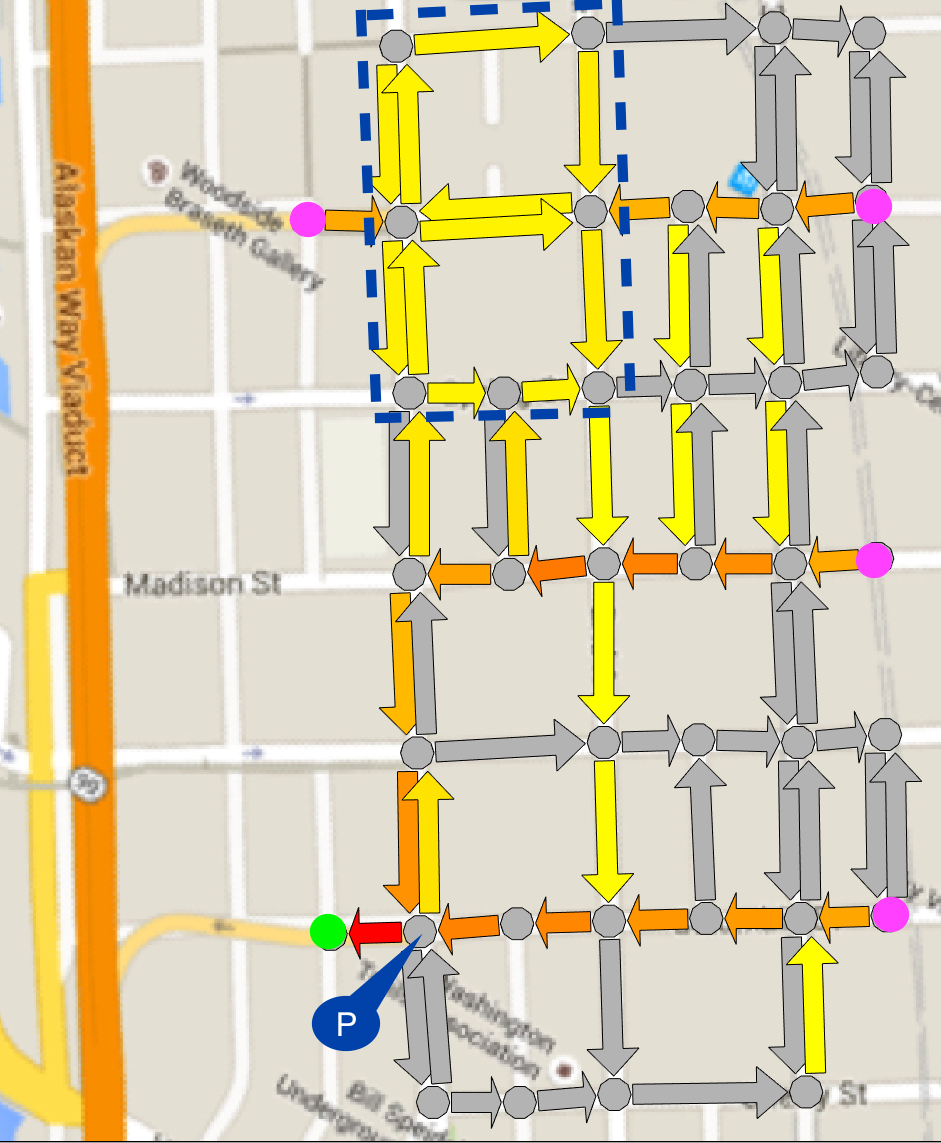}\label{fig:downtown_uncertainty}} \hfill

	\subfloat[Equilibrium inefficiency as a function of $r$]{\includegraphics[width=0.7\columnwidth]{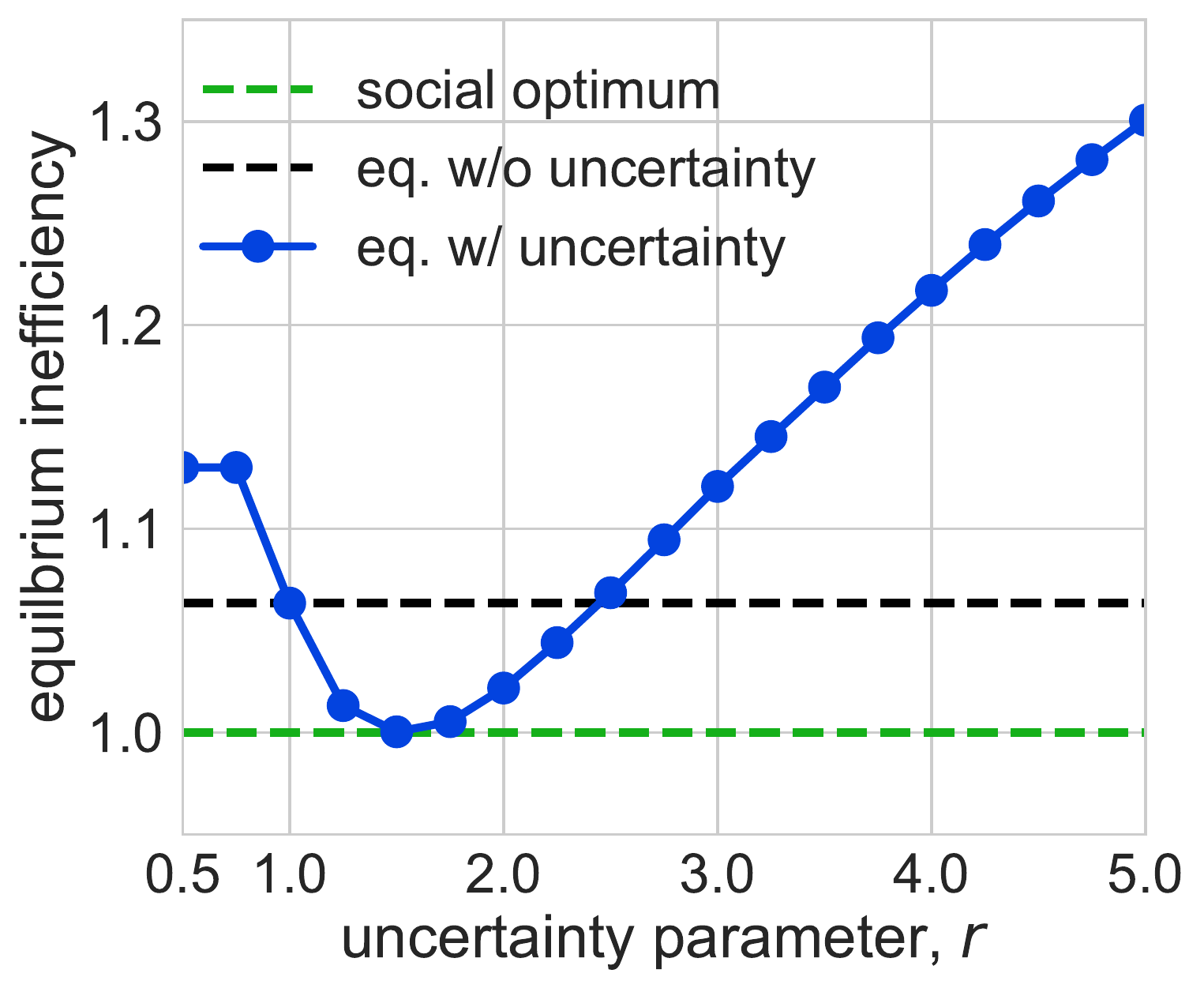}\label{fig:downtown_poa}
	}
	\caption{Setup and results for the parking and through traffic example
		modeled on a region of downtown Seattle.
		In (a) and (b), the network superimposed on the
		corresponding area in downtown Seattle is shown.
		The parking population originates at
		the magenta nodes, whereas the through traffic begins at every grey node
		and terminates at the green node.
		The blue
		 dotted box represents the on-street parking zone and the parking
		symbol (`P') is the location of the off-street parking garage. The color on each edge depicts the intensity of flow on that edge: the
		intensity increases as the color transitions from gray to red.
		}
	\label{fig:downtown}
\end{figure}

In a similar manner to the toy example, we take a real-world urban traffic
network (depicted in Fig.~\ref{fig:downtown}) that captures a slice of a highly congested area in downtown Seattle. The network contains both on-street and off-street
parking options and the parking population experiences both a travel latency cost and a parking cost, both of which we
model as affine functions.  Once again, this can be converted to a standard two-commodity instance by adding a new (fake) destination node $t_{\type_2}$ and including \emph{fake edges} from (i) the top left and bottom right nodes in the on-street parking zone  (boundary nodes on the blue colored dotted area in Fig.~\ref{fig:downtown_nouncertainty}) to $t_{\type_2}$; (ii) the node containing the parking garage (marked with a `P' symbol in Fig.~\ref{fig:downtown_nouncertainty}) to $t_{\type_2}$. By adding fake edges from only specific boundary nodes in the on-street parking area, we are able to capture the added congestion due to the cruising and circling behavior exhibited by users searching for parking spots. As with our previous example, we assume that the through traffic faces no uncertainty $(r_{\type_1}(e) = 1$ for all edges) and the parking users face an uncertainty parameter of $r > 0$ only on the fake edges, which affects their perception of the parking costs.

For the simulations, we assume that the parking traffic originates at a few select nodes in the network
(indicated in magenta in Fig.~\ref{fig:downtown}) and wishes to route their flow to
either an on-street parking slot or a garage. On the other hand, the through
traffic originates at every node in the network and has a single destination,
which represents drivers seeking to leave the downtown area via state highway
$99$. 

The parameters for our simulations were chosen similarly to the example in
Section~\ref{sec:simulations_simple}. Specifically, the cost functions for the
two parking structures were based on an estimation of the number of available
parking slots (which influence the wait times), and the hourly price for parking
for both on-street and garage parking in Seattle. Furthermore, owing to the uniformity of the downtown roads, we assumed that all edges in the network have the same congestion cost function, which were sufficiently scaled in order to ensure that the parking costs are comparable to the transit costs.

As seen in Fig.~\ref{fig:downtown_nouncertainty}, in the downtown Seattle
network, the equilibrium without uncertainty is sub-optimal as more parking
users select the cheaper on-street option. This leads to heavy congestion in the network (indicated by the red edges) as parking users distributed across the network approach the on-street parking area. That being said, the equilibrium without uncertainty is not \emph{as sub-optimal} as the simple network in Fig.~\ref{fig:parkingrouting} as the cost functions are more symmetric and parking users who originate closer to the parking garage prefer using that option despite the higher price. 

In the presence of uncertainty, we observe an interesting phenomenon. When the
parking users are cautious, the garage option becomes more preferable to users
who are equidistant from both parking locations. The parking users distribute
themselves more evenly across the two options, which in turn leads to lesser
congestion in the middle of the network.

It is well-known that sub-optimal behavior by the parking users can cascade and lead to increased congestion for through traffic~\cite{shoup2007gone}. Our simulation results indicate that the routes adopted by the parking users under uncertainty helps alleviate some of this congestion.

Fig.~\ref{fig:downtown_poa} shows the inefficiency of each of the equilibria as a function
of the uncertainty parameter value of the parking users. Specifically, at $r \approx 2$, the
equilibrium solution coincides with the socially optimal flow. On the other
hand, for $r < 1$, the social cost of the equilibrium solution
increases because more users select the on-street parking option. This
in turn leads to heavier congestion in the rest of the network.  Once again, we observe that over-estimation can lead to a decrease in social cost even when $r > 2$ (up to $r \approx 2.5$ in this case). However, unlike our results in Fig.~\ref{fig:parkingpop}, significant cost over-estimation leads to large inefficiency--- this is due to the specific placement of the parking garage in the downtown example. More precisely as $r$ increases, a large fraction of the population favors the garage, whose location intersects with the route adopted by the through traffic. Therefore, beyond a certain threshold, the negative effects of congestion outweigh the positive effects of users choosing the parking garage leading to sub-optimal equilibrium for $r > 2.5$.


We conclude by remarking that even though cautious behavior results in only a small improvement in the
price of anarchy (see Fig.~\ref{fig:downtown_poa}), even a small improvement in daily congestion in downtown areas could result in economic gains.

\section{Conclusions and Future Work}
\label{sec:conclusions}
In this work, we consider a multi-commodity selfish routing game where different
types of users face different levels of uncertainty quantified by a
multiplicative parameter $r_\theta$. Broadly classifying the user attitudes as
cautious and not-cautious, we provide several theoretical results highlighting
the effect that when users over-estimate their network costs, equilibrium
quality tends to improve and vice-versa when users under-estimate the costs. To
our knowledge, this is the first work that provides a characterization of equilibrium quality as a function of uncertainty in networks with multiple types of users, each facing a different level of uncertainty.

Although we focus on linear congestion cost functions, many of our results extend naturally to polynomial cost functions
of the form $C_e(x_e) = a_e{x_e}^d + b_e$. In fact, we showed that uncertainty becomes more favorable as the degree $d$ of the cost function increases. An important direction of future work is to extend the results to  arbitrary convex functions.

\appendix
\label{sec:appendix}
\subsection{Proof of Lemma~\ref{lem_technical_socialcost}}
\label{app:lem:technical_socialcost}
Recall from Proposition~\ref{prop:potential} that the equilibrium solution
$\tilde{\vec{x}}$ minimizes the corresponding potential function
$\Phi_r(\vec{x})$. Therefore, for any arbitrary flow vector $\vec{x}'$, we have
that $\Phi_r(\tilde{\vec{x}}) - \Phi_r(\vec{x}') \leq 0$. Indeed, 
\begin{align}
\text{C}&(\vec{x}')  = \textstyle\sum_{e \in \E}(a_e x'_e + b_e)x'_e \nonumber\\
& =\textstyle\sum_{e \in \E}\big(a_e(x'_e)^2 + b_e\sum_{\type \in
\T}2\frac{x'^\type_e}{r_\type} + b_e\sum_{\type \in \T}(1-\frac{2}{r_\type})
x'^\type_e\big)\nonumber\\
& = \textstyle2\Phi_r(\vec{x}') - \sum_{e \in \E}b_e\sum_{\type \in \T}(\frac{2}{r_\type} - 1)x'^\type_e.  \label{eqn_costtrans}
\end{align}
Applying \eqref{eqn_costtrans} to the solutions $\tilde{\vec{x}}$ and $\vec{x}$, we get that
\begin{align}
\text{C}(\tilde{\vec{x}}) -  \text{C}(\vec{x}) & = 2\Phi_r(\tilde{\vec{x}}) -
2\Phi_r(\vec{x}) \notag\\
&\qquad- \textstyle\sum_{e \in \E}b_e\sum_{\type \in \T}(\frac{2}{r_\type} - 1)(\tilde{x}^\type_e - x^\type_e)\\
& \label{eq:penult}\leq - \textstyle\sum_{e \in \E}b_e\sum_{\type \in \T}(\frac{2}{r_\type} - 1)(\tilde{x}^\type_e -
x^\type_e) \\
& =\textstyle -\sum_{\type \in \T} (\frac{2}{r_\type} - 1)\sum_{e \in \E}b_e \Delta x^{\type}_e.
\end{align}
where the  inequality \eqref{eq:penult} follows from the fact $\Phi_r(\vec{\tilde{x}}) - \Phi_r(\vec{x}) \leq 0$. 
 \hfill$\blacksquare$
\subsection{Proof of Lemma~\ref{lem:lem_technical_poa}}
\label{app:proof:lem_technical_poa}
	Fixing all of the vectors and the function in the left-hand side of
   \begin{equation}\textstyle \frac{f(x)x}{f(x')x' + \sum_{i=1}^n (x_i - x'_i) f(r_i x) } \leq
  4\left(4r_\ast - (r^\ast)^2\right)^{-1},
  \label{eq:poaabstract}
  \end{equation}
    we first derive a lower bound on the denominator
    (i.e.~identify when the denominator is minimized over the space of all valid instantiations of the parameter set). Let us begin with the second term in the denominator:
%
	\begin{align*}
&\textstyle\sum_{i=1}^n (x_i - x'_i) f(r_i x) = \textstyle\sum_{i=1}^n (x_i - x'_i) (r_i ax + b)\\
& \qquad\qquad= \textstyle\sum_{i=1}^n x_ir_i ax - \sum_{i=1}^n x'_i r_i ax  + b (x - x')\\
&  \qquad\qquad\geq \textstyle\sum_{i=1}^n x_ir_{*} ax - \sum_{i=1}^n x'_i r^* ax + b (x - x')\\
&  \qquad\qquad\geq  x r_{*} ax  - x' r^* ax + b (x - x')
	\end{align*}
Using this upper bound on the rest of the terms in the denominator, we get that
\begin{equation*}
\textstyle f(x')x' + \sum_{i=1}^n (x_i - x'_i) f(r_i x) \geq a(x')^2 + bx + xr_{*}ax - x'r^*ax.
\end{equation*}
For any given fixed value of $x$, consider the function $a(x')^2 - x'r^* ax$: by basic calculus, its minimum value is attained if $x' = \frac{r^*x}{2}$. In other words, for any $x,x'$, we can conclude that $a(x')^2 - x'r^*ax \geq (r^*)^2\frac{ax^2}{4} - (r^*)^2\frac{ax^2}{2} = -(r^*)^2\frac{ax^2}{4}$. Substituting this lower bound into the above equation, we get that
\begin{align*}
\textstyle f(x') + \sum_{i=1}^n (x_i - x'_i) f(r_i x) & \textstyle\geq -(r^*)^2\frac{ax^2}{4} + bx + r_*ax^2.
\end{align*}
Now that we have removed the dependence on $x'$, we can substitute this back
into~\eqref{eqn_poaabstract} to get that
\begin{align*}
 \textstyle\frac{f(x)x}{f(x')x' + \sum_{i=1}^n (x_i - x'_i) f(r_i x) } &  \textstyle\leq \frac{ax^2 + bx}{-\frac{(r^*)^2ax^2}{4} + bx + r_*ax^2} \\
& \textstyle\leq \frac{ax^2}{-\frac{(r^*)^2 ax^2}{4} + r_* ax^2}
=\frac{4}{4r_* - (r^*)^2}
\end{align*}
which completes the proof.\hfill$\blacksquare$
\subsection{Proof of Lemma~\ref{lem:flowrerouting}}
\label{app:hetproof}

We prove Lemma~\ref{lem:flowrerouting} via induction. 
Consider the following inductive claim.
\begin{claim} Given instances $\mc{I}=\{G,\mathcal{T}, (s,t),
(\mu_{\type_1}, \mu_{\type_2}),$ $(1,r), (C_e)_{e\in \E}\}$ and $\mc{I}'=
\{G,\mathcal{T}, (s,t), (\mu'_{\type_1}, \mu'_{\type_2}), (1,1),$ $(C_e)_{e\in
\E}\}$
where the graph $G$ is linearly independent, let $\vec{y}$ and $\vec{y}^1$
denote the Nash equilibria of the two instances, respectively. Then,  if
\begin{align}\textstyle\sum_{e \in p}C_e(y_e) \leq \sum_{e \in p}C_e(y^1_e) , \  \forall~p \in
    \mathcal{P} \text{~s.t.~} y^{\type_1}_p > 0,
\label{eqn_inductivecondition}
\end{align}
it must be the case that $y^{\type_1}_p \leq y^1_p$ for all $p \in \mathcal{P}$.
\label{claim:inductive}
\end{claim}
To prove Lemma~\ref{lem:flowrerouting}, it is sufficient to prove the above claim. Indeed,
setting $\mathcal{I} = \mathcal{G}$ and
$\mathcal{I}' = \mathcal{G}^1$, we can directly apply the claim
to prove the lemma statement. In order to verify that the condition from
\eqref{eqn_inductivecondition} is satisfied, first observe that linearly
independent networks are a special case of series-parallel networks. Now,
applying~\cite[Lemma
3]{milchtaich06} to the flows $\vec{y}$ and $\vec{y}^1$ as defined in the statement of Lemma~\ref{lem:flowrerouting}, we get that there exists a path $p^* \in \mathcal{P}$ with $y^1_{p^*} > 0$ such that $\sum_{e \in p^*}C_e(y_e) \leq \sum_{e \in p^*}C_e(y^1_e)$. However, $p^*$ is a min-cost path in $\vec{y}^1$ and its cost in $\vec{y}$ is an upper-bound on that of any min-cost path used by users of type $\type_1$. This immediately implies that for any $p$ with $y^{\type_1}_p > 0$ its cost in $\vec{y}$ (where it is a min-cost path for users of type $\type_1$) must be smaller than or equal to its cost in $\vec{y}^1$ (where all users have no uncertainty).


The proof of Claim~\ref{claim:inductive} proceeds by induction on the number
of edges in the graph $G = (V,\E)$---that is, given a graph $G$, we assume that the
inductive claim holds for all instances defined on any graph $G' = (V',\E')$
such that $|\E'| < |\E|$ and proceed from there.

We also state a property of linearly-independent graphs, which will be useful for our proof.
\begin{proposition}[{\cite[Proposition 5]{milchtaich06}}] A two terminal network $G$ is linearly-independent if and only if:
(i) it consists of a single edge, or (ii) it is the result of connecting two linearly-independent networks in parallel,
		or (iii) it is the result of connecting in series a linearly independent network and a network with a single edge.
	\label{prop_sp}
\end{proposition}

\begin{IEEEproof}[Proof of Claim~\ref{claim:inductive}]
Let us start with the base case. The inductive hypothesis is trivially true when
the graph $G$ consists of a single edge between the source and sink (i.e.~$\E =
\{e \}$). All of the users must route their flow on this edge only, regardless of their uncertainty level. Since we are given (by assumption) that the instance satisfies the condition that
$$C_e(y_e) \leq C_e(y^1_e),$$
by monotonicity of the cost functions, we get that $y_e \leq y^1_e$. Of course,
this implies that $y^{\type_1}_e \leq y^1_e$ and the inductive claim thus
holds for the base case.

Now, let us consider the inductive case. Consider the inductive claim with
respect to an arbitrary linearly independent graph $G = (V,\E)$ consisting of
two or more edges, and assume that the inductive hypothesis holds for all
instances defined on linearly independent graphs $G'=(V',\E')$ such that $|\E'|
< |\E|$. From Proposition~\ref{prop_sp}, we know that $G$ consists of two
sub-graphs $G_1 = (V_1, \E_1)$ and $G_2 = (V_2, \E_2)$ connected either in
parallel or in series with $|\E_2| = 1$ (i.e., $G_2$ is simply a single edge).
We consider both cases, apply the inductive claim recursively to both
sub-graphs, and merge the resulting flows to prove the inductive claim for the original graph $G$.

Let $(s_1, t_1)$, $(s_2, t_2)$ denote the origin-destination pairs for $G_1$ and
$G_2$ respectively. We use $\mathcal{P}_1$ to denote the set of $s_1$-$t_1$
paths in $G_1$ and $\mathcal{P}_2$ to denote the set of $s_2$-$t_2$ paths in
$G_2$. When $G_1$ and $G_2$ are connected in series (case 1), we have that $s_2
= t_1$. When they are connected in parallel (case 2), $s_1 = s_2, t_1 = t_2$. 
We treat each of these cases separately.

(case 1) \emph{Subgraphs connected in series}. 
We first introduce some additional notation required for this part of the proof. As we did in the proof of Theorem~\ref{thm_characterization}, we use $\vec{y}(1)$ and $\vec{y}(2)$ to denote the sub-flows of $\vec{y}$  with respect to graphs $G_1$ and $G_2$. Similarly, we define $\vec{y}^1(1)$ and $\vec{y}^1(2)$ to be the sub-flows of $\vec{y}^1$.  Recall that $G_2$ consists simply of a single edge. 

Since $\vec{y}$ is an equilibrium for $\mathcal{I}$, it must be the case that $\vec{y}(1)$ and $\vec{y}(2)$ are equilibria for $G_1, G_2$ respectively for suitably defined sub-instances, 
$\mc{I}_1=\{G_1,\mathcal{T},$ $(s_1,t_1), (\mu_{\type_1},$ $\mu_{\type_2}), (1,r),
(C_e)_{e\in \E_1}\}$
and $\mc{I}_2=\{G_2,\mathcal{T},$ $(s_2,t_2),
(\mu_{\type_1},\mu_{\type_2}),$ $
(1,r), (C_e)_{e\in \E_2}\}$,
respectively. In a similar manner,  $\vec{y}^1(1)$ and $\vec{y}^1(2)$ represent equilibria for the instances 
$\mc{I}'_1=\{G_1,\mathcal{T},$ $(s_1,t_1), (\mu'_{\type_1}, \mu'_{\type_2}),$
$(1,1), (C_e)_{e\in \E_1}\}$
and 
$\mc{I}'_2=\{G_2,\mathcal{T},$ $(s_2,t_2), (\mu'_{\type_1}, \mu'_{\type_2}),
(1,1), (C_e)_{e\in \E_2}\}$, respectively. These are formally summarized in the notation table below.


\begin{table}[h]
	\centering
	\caption{Notation Table providing different instance definitions and identifying the Nash equilibrium (N.E.) for each sub-instance when $G_1$ and $G_2$ are connected in series.}
	\begin{tabular}{|l|l|c|}
		\hline
		 & Game Instance Definition & N.E. \\
		\hline
		$\mathcal{I}$ & $\{G,\mathcal{T}, (s,t), (\mu_{\type_1}, \mu_{\type_2}), (1,r), (C_e)_{e\in \E}\}$ &  $\vec{y}$\\
		$\mathcal{I}_1$ & $\{G_1,\mathcal{T}, (s_1,t_1), (\mu_{\type_1}, \mu_{\type_2}), (1,r), (C_e)_{e\in \E_1}\}$ & $\vec{y}(1)$ \\
		$\mathcal{I}_2$ & $\{G_2,\mathcal{T}, (s_2,t_2), (\mu_{\type_1}, \mu_{\type_2}), (1,r), (C_e)_{e\in \E_2}\}$ & $\vec{y}(2)$\\
		$\mathcal{I}'$ & $\{G,\mathcal{T}, (s,t), (\mu'_{\type_1}, \mu'_{\type_2}), (1,1), (C_e)_{e\in \E}\}$ & $\vec{y}^1$ \\
		$\mathcal{I}'_1$ & $\{G_1,\mathcal{T}, (s_1,t_1), (\mu'_{\type_1}, \mu'_{\type_2}), (1,1), (C_e)_{e\in \E_1}\}$ & $\vec{y}^1(1)$  \\
		$\mathcal{I}'_2$ & $\{G_2,\mathcal{T}, (s_2,t_2), (\mu'_{\type_1}, \mu'_{\type_2}), (1,1), (C_e)_{e\in \E_2}\}$ & $\vec{y}^1(2)$ \\
		\hline
	\end{tabular}
	\label{table_notation_series}
\end{table}

Proceeding with the proof, we know from \eqref{eqn_inductivecondition} that for
any path $p \in \mc{P}$ with $y^{\type_1}_p > 0$, $\sum_{e \in p}C_e(y_e) \leq\sum_{e \in p}C_e(y^1_e)$. Dividing the path $p$ into sub-paths $p_1$ and $p_2$ representing its intersection with $G_1, G_2$ respectively, we get that
\begin{align*}
&\textstyle\sum_{e \in p_1}C_e(y_e) + \sum_{e \in p_2}C_e(y_e) \notag\\
&\quad\leq \textstyle\sum_{e \in p_1}C_e(y_e^1) + \sum_{e \in p_2}C_e(y_e^1).
\end{align*}
Based on the above inequality, we infer one of the following two cases must be
true: \begin{equation}
\textstyle\sum_{e \in p_1}C_e(y_e) \leq \sum_{e \in p_1}C_e(y^1_e)
\label{eq:cond1}
\end{equation}
or
\begin{equation}
    \textstyle\sum_{e
\in p_2}C_e(y_e) \leq \sum_{e \in p_2}C_e(y^1_e).
\label{eq:cond2}
\end{equation}
Separately, we prove the inductive claim for each case.

Suppose that \eqref{eq:cond1} holds.
Since the graph $G_1$ is linearly independent, we can apply the inductive claim to instances $\mathcal{I}_1$ and $\mathcal{I}'_1$ to get that $y(1)^{\type_1}_p \leq y^1(1)_p$ for every path $p \in \mathcal{P}_1$. However, recall that $G_2$ consists of only one edge (call it $e_2$). This implies that there is a one-to-one correspondence between every path in $\mathcal{P}$ and path in $\mathcal{P}_1$. Specifically, for every path $p \in \mathcal{P}$, there exists a unique path $p_1 \in \mathcal{P}_1$ such that $p = p_1 \cup \{e_2\}$. Since the mapping is a bijection, this implies that for every $p = p_1 \cup \{e_2\}$, it must be the case that $y^{\type_1}_p = y(1)^{\type_1}_{p_1}$ and $y^1_p = y^1(1)_{p_1}$. Informally, the addition of a single common edge to all paths does not alter the flow distribution. Therefore, we can conclude that for every path $p \in \mathcal{P}$,
\[y^{\type_1}_{p}  = y(1)^{\type_1}_{p_1}  \leq y^1(1)_{p_1} = y^1_p.\]
This proves the inductive claim for this case. 

Next,  suppose that \eqref{eq:cond2} holds. 
Once again, since $G_2$ consists of a single edge $e_2$, the condition implies that $y_{e_2} = y(2)_{e_2} \leq y^1_{e_2} = y^1(2)_{e_2}$. Thus, we infer that conditional upon   $\sum_{e \in p_2}C_e(y_e) \leq \sum_{e \in p_2}C_e(y^1_e)$, $\mu_{\type_1} + \mu_{\type_2} \leq \mu'_{\type_1} + \mu'_{\type_2}$. That is, the total population for instance $\mathcal{I}$ is smaller than or equal to the total population for instance $\mathcal{I}'$.

Since linearly independent networks are a special case of series-parallel, we
can now apply~\cite[Lemma~3]{milchtaich06} for flows $\vec{y}(1)$ and $\vec{y}^1(1)$ on graph $G_1$. Since $\mu_{\type_1} + \mu_{\type_2} \leq \mu'_{\type_1} + \mu'_{\type_2}$, this of course, implies that there exists at least one origin-destination path $p^*$ in $G_1$ with $y^1(1)_{p^*} > 0$ satisfying $\sum_{e \in p^*}C_e(y(1)_e) \leq \sum_{e \in p^*}C_e(y^1(1)_e)$. Therefore, we can conclude that for any path $p \in \mathcal{P}_1$ with $y^{\type_1}_p > 0$,  we have,
\begin{equation}
\textstyle \sum_{e \in p}C_e(y(1)_e) \leq \sum_{e \in p}C_e(y^1(1)_e).
\end{equation}

The above equation symbolizes the induction condition as described in \eqref{eqn_inductivecondition}. Applying the inductive hypothesis recursively to instances $\mathcal{I}_1, \mathcal{I}'_1$, and using the same reasoning as before, we conclude that for every $s$-$t$ path $p \in \mathcal{P}$, $y^{\type_1}_p \leq y^1_p$.  This concludes the proof for the series case.

(case 2) \emph{Subgraphs $G_1, G_2$ connected in parallel to obtain $G$.}
The proof for this case proceeds similarly to the case where the networks are connected in series. Once again, we divide the flows $\vec{y}$ and $\vec{y}^1$ into sub-flows $\vec{y}(1), \vec{y}(2)$ and $\vec{y}^1(1)$ and $\vec{y}^1(2)$. Moreover, we use $\mu_{\type_1}(1), \mu_{\type_2}(1)$ to denote the total flow without and with uncertainty in sub-graph $G_1$ and similarly so, for all the other subgraphs and sub-flows.  A comprehensive notation table is listed below.

\begin{table}[h]
	\centering
	\caption{Notation table providing different instance definitions and identifying the Nash equilibrium (N.E.) for each sub-instance when $G_1$ and $G_2$ are connected in parallel.}
	\begin{tabular}{|l|l|l|}
		\hline
		 & Game Instance Definition & N.E. \\
		\hline
		$\mathcal{I}$ & $\{G,\mathcal{T}, (s,t), (\mu_{\type_1}, \mu_{\type_2}), (1,r), (C_e)_{e\in \E}\}$ &  $\vec{y}$\\
		$\mathcal{I}_1$ & $\{G_1,\mathcal{T}, (s,t), (\mu_{\type_1}(1), \mu_{\type_2}(1), (1,r), (C_e)_{e\in \E_1}\}$ & $\vec{y}(1)$ \\
		$\mathcal{I}_2$ & $\{G_2,\mathcal{T}, (s,t), (\mu_{\type_1}(2), \mu_{\type_2}(2)), (1,r), (C_e)_{e\in \E_2}\}$ & $\vec{y}(2)$\\
		$\mathcal{I}'$ & $\{G,\mathcal{T}, (s,t), (\mu'_{\type_1}, \mu'_{\type_2}), (1,1), (C_e)_{e\in \E}\}$ & $\vec{y}^1$ \\
		$\mathcal{I}'_1$ & $\{G_1,\mathcal{T}, (s,t), (\mu'_{\type_1}(1), \mu'_{\type_2}(1), (1,1), (C_e)_{e\in \E_1}\}$ & $\vec{y}^1(1)$  \\
		$\mathcal{I}'_2$ & $\{G_2,\mathcal{T}, (s,t), (\mu'_{\type_1}(2), \mu'_{\type_2}(2), (1,1), (C_e)_{e\in \E_2}\}$ & $\vec{y}^1(2)$ \\
		\hline
	\end{tabular}
	\label{table_notation_parallel}
\end{table}

 Before applying the inductive claim recursively to the two subgraphs, we need
 to verify that the condition specified in \eqref{eqn_inductivecondition} is
 satisfied for the subgraphs. We know that the instances $\mathcal{I},
 \mathcal{I}'$ satisfy the condition in \eqref{eqn_inductivecondition} (this is
 part of our assumption). That is, for any $p \in \mathcal{P}$ such that $y^{\type_1}_p > 0$
\begin{equation}
\label{eqn_inductiveparallel}
\textstyle \sum_{e \in p}C_e(y_e) \leq \sum_{e \in p}C_e(y^1_e).
\end{equation}
Since the graphs $G_1, G_2$ are connected in parallel, the set of edges and
paths in each graph are disjoint. In other words, $\mathcal{P} = \mathcal{P}_1
\cup \mathcal{P}_2$. Moreover, the path-flows in $\vec{y}$ coincide with those
in $\vec{y}(1)$ or $\vec{y}(2)$---e.g., for any $p_1 \in \mathcal{P}_1$, we have
that $y_{p_1} = y(1)_{p_1}$ and so on.  This combined with
\eqref{eqn_inductiveparallel} implies that for any path $p_1 \in \mathcal{P}_1$
with $y(1)^{\type_1}_{p_1} > 0$, $\sum_{e \in p_1}C_e(y(1)_e) \leq \sum_{e \in
p_1}C_e(y^1(1)_e)$. Therefore, the flows $\vec{y}(1)$ and $\vec{y}^1(1)$ satisfy
the inductive condition. Similarly, we can show
that the flows  $\vec{y}(2)$ and $\vec{y}^1(2)$ also satisfy \eqref{eqn_inductivecondition}.



Applying the inductive claim to the instances $\mathcal{I}_1$ and $\mathcal{I}'_1$, we get that
\begin{equation}
	\label{eqn_inductiveparallel1}
	y(1)^{\type_1}_{p_1} \leq y^1(1)_{p_1} ~~ \forall p_1 \in \mathcal{P}_1
\end{equation}
Similarly, we can apply the inductive claim to instance pair $\mathcal{I}_2, \mathcal{I}'_2$ to get that
\begin{equation}
\label{eqn_inductiveparallel2}
y(2)^{\type_1}_{p_2} \leq y^1(2)_{p_2} ~~ \forall p_2 \in \mathcal{P}_2
\end{equation}
Since $\vec{y} = \vec{y}(1) \cup \vec{y}(2)$ and $\vec{y}^1 =
\vec{y}^1(1) \cup \vec{y}^1(2)$, the inductive claim follows immediately. That is, merging the statements in \eqref{eqn_inductiveparallel1} and~\eqref{eqn_inductiveparallel2}, we get that for all $p \in \mathcal{P}$, the inequality $y^{\type_1}_p \leq y^1_p$ holds.
This completes the proof. 
\end{IEEEproof}
\subsection{Technical Claims}
\label{app:techclaims}

\begin{lemma}
	\label{lem:techconvex}
	Given any convex, continuously differentiable, non-decreasing function $f(x)$, for any $x_1, x_2 \geq 0$ in its domain, we have that
	\begin{equation}
	f(x_2) - f(x_1) \leq f'(x_2)(x_2 - x_1),
	\end{equation}
    where $f'(x_2)$ is the derivative of $f(x)$ evaluated at $x_2$.
	\end{lemma}
	\begin{IEEEproof}
		We prove the result in two cases. First, suppose that $x_2 \geq x_1$. Since $f(x)$ is convex, non-decreasing, it must mean that its derivative $f'(x)$ is also non-negative and non-decreasing, and therefore, $f'(x_2) \geq f'(x)$ for all $x \in [x_1, x_2]$ Therefore, we have that
	\[\textstyle f(x_2) - f(x_1) \leq \int_{x=x_1}^{x_2} f'(x_2) dx =
    f'(x_2)(x_2 - x_1).\]

		Next, consider the case when $x_2 \leq x_1$. In this case, $f(x_2) - f(x_1)$ is negative or zero. So, similar to the above case we have that
		\[\textstyle f(x_1) - f(x_2) \geq \int_{x=x_1}^{x_2} f'(x_2) dx =
        f'(x_2)(x_1 - x_2).\]
		Reversing the above inequality gives us the lemma. 		\end{IEEEproof}

\bibliographystyle{IEEEtran}
\bibliography{2018acc_parkingrefs}

\end{document}